\documentclass[%
 reprint,
 amsmath,amssymb,
 aps,
 prc,
floatfix,
]{revtex4-2}
\usepackage{longtable}

\usepackage{changes} 
\usepackage[para,online,flushleft]{threeparttable}
\usepackage{graphicx}  
\usepackage{dcolumn}   
\usepackage{bm}        
\usepackage{amssymb}   
\usepackage{amssymb,amsmath,amsfonts} 
\usepackage{braket}    
\usepackage{booktabs}
\usepackage{multirow}
\usepackage{adjustbox}
\usepackage{gnuplottex}
\usepackage[T1]{fontenc}
\usepackage[utf8]{inputenc}
\usepackage{color}
\usepackage{dsfont}
\usepackage{tikz}
\usepackage{pgfplots}
\usepgfplotslibrary{fillbetween}
\usetikzlibrary {arrows.meta}
\usepackage[section]{placeins}
\usepackage{changes}

\usepgfplotslibrary{groupplots}
\pgfplotsset{compat=newest}
\usepgfplotslibrary{units}
\usetikzlibrary{patterns}

\newcommand{\msun}{M_{\odot}}

\begin{document}
	

    %
\title{Universal Relation for the Neutron Star Maximum Mass within Relativistic Mean-Field Theories}
\author{Gihwan Nam} 
\email{namgh@yonsei.ac.kr}
\affiliation{Department of Physics, Yonsei University, Seoul 03722, South Korea}
\author{Yeunhwan Lim}
\email{ylim@yonsei.ac.kr}
\affiliation{Department of Physics, Yonsei University, Seoul 03722, South Korea}
\author{Jeremy~W. Holt}
 \email{holt@physics.tamu.edu}
\affiliation{Cyclotron Institute, Texas A\&M University, College Station, TX 77843, USA}
\affiliation{Department of Physics and Astronomy, Texas A\&M University, College Station, TX 77843, USA }
\date{\today}
	
\begin{abstract} 
We obtain a universal relation 
for the neutron star maximum mass arising from a particular combination of the saturation density ($n_0$), the effective mass ($m^*$), and (when present) the vector meson self-coupling constant ($\zeta$) within the relativistic mean-field model framework.
Observations of massive neutron stars heavier than $\sim 2M_{\odot}$ have 
eliminated the softest equation of state from consideration and impose strong constraints on nuclear interactions used to model dense nuclear matter. To date there have been numerous attempts to refine relativistic mean-field models by including the presence of additional mesons, such as the delta meson, and couplings. 
We show that current RMF models, including our own constructions, exhibit a maximum neutron star mass that is primarily determined by the combination of the saturation density, the effective mass at saturation, and the vector meson self-coupling constant. When constraining the pure neutron matter equation of state using chiral effective field theory (ChEFT) at low densities, 250 parameter sets were generated to derive an empirical formula for the maximum mass of neutron stars and apply the formula with the present relativistic mean field models.
\end{abstract}
\maketitle

\section{Introduction}
Understanding the physics of ultra-dense matter in neutron stars remains a challenge in nuclear physics due to the limitations of terrestrial experiments
in exploring highly neutron-rich matter as well as matter beyond about twice normal nuclear densities. 
Progress on these fronts has been achieved through heavy-ion collisions (HIC), but at present there is not enough quality data to significantly constrain the uncertainties for isospin-asymmetric or neutron-rich nuclear matter\,\cite{Danielewicz2002,Huth2021,Most2022,Duguet2025}. 
Neutron star observations, on the other hand, offer a unique opportunity to investigate the properties of dense nuclear matter at supranuclear densities and extreme isospin asymmetry. Macroscopic observables, such as the neutron star mass, radius, and tidal deformability, can provide invaluable constraints for theoretical nuclear models.
Composed primarily of neutrons, along with a small fraction of protons, electrons, and possibly exotic particles, neutron stars are the densest observable objects in the universe. Their core density may reach up to ten times the saturation density of nuclear matter, leading to extreme physical conditions characterized by strong nuclear interactions and significant relativistic effects at the highest densities. These features make neutron stars an essential testing ground for the physics of dense matter beyond what can be achieved in terrestrial laboratories.\,\cite{Lattimer2004}.

One of the most striking general relativistic effects in neutron stars is the existence of a maximum mass — a critical threshold beyond which no stable configuration can exist. The predicted maximum mass of a neutron star depends sensitively on the underlying model of dense matter. Since observations of massive neutron stars provide a lower bound on the maximum mass, any model that predicts a maximum mass below the observed value can be considered inconsistent with the correct description of dense matter physics.

A neutron star is composed of several distinct regions. The outermost layer is the atmosphere, consisting of a thin plasma that emits thermal radiation \,\cite{Lattimer2004}. Beneath the atmosphere lies the crust, which is further divided into the outer crust, composed of nuclei and free electrons, and the inner crust, where neutron-rich nuclei coexist with a superfluid neutron background. Deeper inside is the core, which accounts for most of the star's mass. The core is generally divided into the outer core and the inner core, and the outer core is believed to be composed of homogeneous nuclear matter and leptons to maintain charge neutrality. The composition of the inner core remains uncertain and may involve exotic states of matter, such as hyperons, meson condensates, or deconfined quark matter at extremely high densities. However, even the very existence of the inner core is uncertain, primarily because observations of massive neutron stars around $2\,M_{\odot}$ place stringent constraints on the equation of state, which relates pressure to energy density, disfavoring many exotic phases that would otherwise soften it\,\cite{Lattimer2021}.
Nuclear interaction models are employed to obtain the equation of state of homogeneous nuclear matter in the core region of neutron stars.
This is the very reason why neutron stars are regarded as key astrophysical systems for testing theories of dense matter based on nuclear interactions.

Relativistic Mean Field (RMF) theory provides a compact and efficient framework for describing both the structure of finite nuclei and the equation of state of nuclear matter. In particular, RMF theory is particularly advantageous for modeling the equation of state of neutron stars, because fundamental constraints such as causality are automatically satisfied within its framework. Coupling constants in RMF models are typically fine-tuned to accurately reproduce essential nuclear observables, such as binding energies and charge radii of closed-shell nuclei. This approach has demonstrated considerable success in accurately describing experimental data especially for stable nuclei.

The foundation of the Relativistic Mean Field (RMF) theory was established in 1974 by Walecka, who introduced a relativistic field-theoretical approach to nuclear many-body problems\,\cite{walecka1974}. His model, often referred to as the $\sigma-\omega$ model, described nucleons as Dirac particles interacting via the exchange of scalar $\sigma$ and vector $\omega$ mesons. The scalar meson provided attraction, while the vector meson contributed repulsion, leading to a self-consistent description of nuclear saturation. Although the original Walecka model successfully captured the qualitative features of nuclear matter, it suffered from an unrealistic nuclear incompressibility, predicting an equation of state that was too stiff. To overcome the limitations of the Walecka model, Boguta and Bodmer\,\cite{boguta1977} introduced nonlinear self-interactions of the scalar meson. This modification softened the equation of state and improved the description of finite nuclei, making the RMF models more realistic. The inclusion of nonlinear terms was essential to reduce the overbinding problem and better reproduce empirical nuclear properties, such as charge radii. These advancements formed the basis for the development of modern RMF parameterizations.

Throughout the 1980s and 1990s, RMF theory became one of the standard frameworks for exploring the properties of dense matter, with numerous parameter sets developed to refine the description of various nuclear properties\,\cite{Gambhir1990}. Among these, the NL3 parametrization became one of the most widely used models due to its accurate description of finite nuclei properties\,\cite{Lalazissis1997}.
Even though NL3 describes the static properties of finite nuclei quite well,
it does not adequately explain their dynamic properties, such as the isoscalar giant monopole resonance (ISGMR) and the isovector giant dipole resonance (IVGDR).

To resolve the discrepancy between the NL3 model and experimental values of the ISGMR of $^{90} \mathrm{Zr}$  and the  IVGDR of $^{208} \mathrm{Pb}$, 
models incorporating omega meson self-coupling and meson-meson mixing, such as the FSU and IU-FSU models, have been proposed\,\cite{Vretenar2003,Todd-Rutel2005,Fattoyev2010}. 
These models have proven effective in describing key nuclear observables, including the ISGMR of $^{90} \mathrm{Zr}$ and the IVGDR of $^{208} \mathrm{Pb}$, thereby improving the overall description of finite nuclei. However, the maximum mass of neutron stars predicted by the FSU model is too low to be consistent with recent observational data\,\cite{Antoniadis2013,Fonseca2021,Linares2018}. Similarly, the IU-FSU model also fails to predict a sufficiently high maximum mass, suggesting that further refinement of the equation of state for neutron stars is necessary. In response, efforts have been recently made to enhance the maximum mass of neutron stars by including the delta meson $\delta$ in RMF models\,\cite{VirenderThakur2022,FanLi2022,Miyatsu2023}. 
Unfortunately, the effective mass splitting induced by the delta meson exhibits an opposite sign compared to the non-relativistic case and experiments, leading to ambiguities in its physical interpretation\,\cite{BaoanLi2018}. 
Consequently, it remains uncertain up to what maximum mass RMF theory can reliably predict neutron stars without compromising the description of finite nuclei. A systematic understanding of the relationship between RMF theory and the maximum mass of neutron stars is required. In this study, we investigate an empirical formula for the maximum mass of neutron stars within the framework of relativistic mean-field (RMF) theory.

This paper is organized as follows. In Section\,\ref{sec:formalism}, we briefly explain the RMF models and our fitting procedure for determining a new parameter set. Many-body perturbation theory with the chiral effective potential is primarily used to fit the pure neutron matter equation of state. Section\,\ref{sec:results} presents the numerical results for the maximum mass of neutron stars and demonstrates how a universal relation can be obtained. In Section\,\ref{sec:conclusion}, we summarize our findings and discuss directions for further analysis.

\section{Relativistic Mean Field Models}\label{sec:formalism}
The formulation of the effective Lagrangian density and the corresponding energy density functionals for the extended RMF model has been extensively discussed in Refs\,\cite{Glendenning2000,Dutra2014,lim17h}. The effective Lagrangian density adopted in this study is given by

\begin{align}
\mathcal{L}
&= \bar\psi\left[\gamma^\mu\bigl(i\partial_\mu - g_vW_\mu -\frac{g_\rho}{2}\bm{\tau}\cdot \bm{b_\mu}\bigr) -(M_N -g_s\phi)\right]\psi   \nonumber \\ 
&
   + \tfrac12\,\partial_\mu\phi\,\partial^\mu\phi
   - \tfrac12\,m_s^2\,\phi^2
   - \tfrac{\kappa}{3}(g_s \phi)^3 - \tfrac{\lambda}{4}(g_s\phi)^4 \nonumber \\ 
& 
   - \tfrac14\,W_{\mu\nu}\,W^{\mu\nu}
   + \tfrac12\,m_v^2\,W_\mu\,W^\mu+ \tfrac{\zeta}{4 }(g_v^2\,W_\mu W^\mu)^2  \\
& 
   - \tfrac14\,\bm{b_{\mu\nu}} \cdot \bm{b^{\mu\nu}} 
   + \tfrac12\,m_\rho^2\,\bm{b_\mu} \cdot \bm{b^\mu} \nonumber\\
&
+ \Lambda_v (g_v^2\,W_\mu W^\mu)
     (g_\rho^2 \bm{b_\mu} \cdot \bm{b^\mu}) \nonumber  
    + \Lambda_{s1}(g_s \phi)(g_\rho^2\,\bm{b_\mu} \cdot \bm{b^\mu}) \\
& + \Lambda_{s2}(g_s^2 \phi^2)(g_\rho^2\,\bm{b_\mu} \cdot \bm{b^\mu}) \nonumber
\end{align}
 where $\psi$ represents an iso-doublet nucleon field that interacts through the exchange of two iso-scalar mesons, the scalar $\sigma$ meson ($\phi$) and the vector $\omega$ meson ($W_\mu$) as well as one iso-vector meson, the $\rho$ meson ($\bm{b}_\mu$). 
 In addition to meson-nucleon interactions, the Lagrangian density also includes scalar and vector self-interactions, as well as meson-meson mixing. 
 The presence of $\omega$ meson self-coupling ($\zeta$) provides additional softening to the equation of state and the meson-meson couplings ($\Lambda_v,\Lambda_{s1},\Lambda_{s2}$) allow the symmetry properties to be adjusted without affecting the symmetric nuclear matter properties. The additional meson-meson coupling between $\sigma$ and $\omega$ could be included. Since their effect is redundant with $\zeta$, they are not included in this study.
 The nucleon mass is denoted as $M_N$, while the meson masses of the sigma, omega, and rho mesons are given as $m_s$, $m_v$, and $m_\rho$ respectively. In this study, we adopt the following representative values: $M_N=939$\,MeV for the nucleon, and $m_s=508.194$\,MeV, $m_v=762.5$\,MeV, and $m_\rho=763$\,MeV for the mesons.
 In the mean field approximation, 
 the meson fields are treated as classical fields:
\begin{equation*}
\langle g_s\phi \rangle = \Phi,\quad
\langle g_v W_\mu \rangle = V_0,\quad
\langle g_\rho \bm{b_\mu} \rangle = B_0.    
\end{equation*}
Combined with the Euler-Lagrange equations, this leads to the following field equations for homogeneous nucleonic matter:
\begin{align}
  &  \bigl[i\gamma^\mu \partial_\mu  -\gamma^0 \bigl(V_0 - B_0\tfrac{\tau_3}{2} \bigr)
      - \bigl(M_N - g_s\,\phi\bigr)\bigr]\psi = 0\\
 &\tfrac{m_s^2}{g_s^2}\,\Phi+ \kappa\,\Phi^2
  + \lambda\,\Phi^3 +\Lambda_{s1} B_0^2  +2\,\Lambda_{s2}  \Phi  B_0^2  = n_s  \label{Eq euler Lagrange:sigma}
\\
 &\tfrac{m_v^2}{g_v^2}\,V_0+ \zeta\,V_0^3 + 2\,\Lambda_v\,V_0\,B_0^2  = \,n \label{Eq euler Lagrange:omega}
  \\
 &\tfrac{m_\rho^2}{g_\rho^2}\,B_0+2\,\Lambda_v\,V_0^2\,B_0+2\Lambda_{s1}\,\Phi B_0 +2\Lambda_{s2}\,\Phi^2 B_0  = \tfrac{n_3}{2} 
\end{align}
where nucleon scalar, vector, and isospin excess densities are denoted by
\begin{equation}
n_s = \langle \bar\psi \psi \rangle,\quad
n   = \langle \psi^\dagger \psi \rangle,\quad
n_3 = \langle \psi^\dagger \,\tau_3\,\psi \rangle.    
\end{equation}
With the energy–momentum tensor $T^{\mu\nu}$ in hand, the energy density $\varepsilon$ and pressure $P$ can be expressed as
\begin{equation}
    \begin{aligned}
    \varepsilon & = \tfrac12\tfrac{m_s^2}{g_s^2}\,\Phi^2
             + \tfrac{\kappa}{3} \,\Phi^3
             + \tfrac{\lambda}{4}\,\Phi^4 
           - \tfrac12\tfrac{m_v^2}{g_v^2}\,V_0^2- \tfrac{\zeta}{4}\,V_0^4 + V_0\,n
             \\
&       -\tfrac12\tfrac{m_\rho^2}{g_\rho^2}\,B_0^2
            + B_0\tfrac{n_3}{2}
-\Lambda_v\,V_0^2\,B_0^2  -\Lambda_{s1}\Phi \, B_0^2 -\Lambda_{s2}\Phi^2  B_0^2\\
&\,\, +\tfrac{1}{\pi^2}\sum_{i=n,p}\int_0^{k_i} k^2 \sqrt{k^2 + {M^*}^2}\,dk \label{Eq energy density} \\
\end{aligned}
\end{equation}
\begin{equation}
    \begin{aligned}
P           &=  - \tfrac12\,\tfrac{m_s^2}{g_s^2} \Phi^2
             - \tfrac{\kappa}{3}\,\Phi^3
             - \tfrac{\lambda}{4}\,\Phi^4
             + \tfrac12\tfrac{m_v^2}{g_v^2}\,V_0^2
             + \tfrac{\zeta}{4}\,V_0^4 \\
&
             + \tfrac12\tfrac{m_\rho^2}{g_\rho^2}\,B_0^2
             + \Lambda_v\,V_0^2\,B_0^2\,+\Lambda_{s1}\Phi \, B_0^2 +\Lambda_{s2}\Phi^2 B_0^2\\
& +\tfrac{1}{3\pi^2}\sum_{i=n,p}\int_0^{k_i}\frac{k^4}{\sqrt{k^2 + {M^*}^2}}\,dk \\
    \end{aligned}
\end{equation}
where the Dirac effective mass is defined by
\begin{equation}
\label{Eq effective mass}
M^* = M_N - \Phi.    
\end{equation}
The saturation density\,($n_0$) is determined by the condition
\begin{equation}
P(n_0) = 0,    
\end{equation}
with vanishing isospin density\, $n_3= 0$.  The dimensionless effective mass at saturation is expressed as
\begin{equation}
m^* = \frac{M^*(n_0)}{M_N},    
\end{equation}
and hereafter we refer to $m^*$ simply as the {\it effective mass}.  The binding energy per nucleon of symmetric nuclear matter ($n_3=0$) at the saturation density is given as
\begin{equation}
\frac{B}{A} = \frac{\varepsilon(n_0)}{n_0} - M_N ,    
\end{equation}
and the incompressibility is given by
\begin{equation}
\label{Eq incompressibility}
K = 9 \left.\frac{\partial P}{\partial n}\right|_{n=n_0,\;n_3=0}.    
\end{equation}

\begin{figure}
    \centering
    \includegraphics[width=1.0\linewidth]{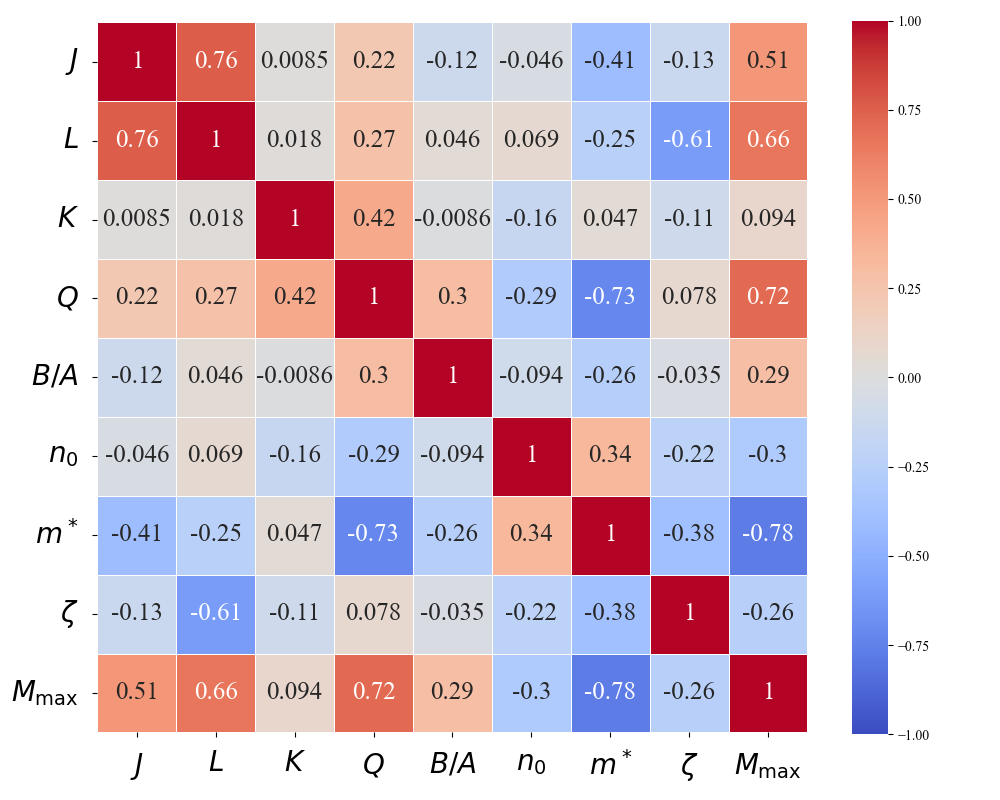}
     \caption{
        Pearson correlation analysis between nuclear matter properties and the maximum mass of a neutron star. 
        The nuclear matter properties include the symmetry energy ($J$), its slope ($L$), incompressibility ($K$), skewness ($Q$), binding energy per nucleon ($B/A$), saturation density ($n_0$), effective mass ($m^*$), and the $\zeta$ coupling constant. 
        The correlation coefficients are computed based on the parameter sets listed in Table~\ref{M2.35 nuclear matter}.
    }
    \label{fig:cor}
\end{figure}
  As a first step, we performed a correlation analysis to gain insight from the existing parameter sets and to identify which nuclear matter properties contribute most significantly to determining the maximum mass ($M_{\mathrm{max}}$) of a neutron star (Figure \ref{fig:cor}). The parameter sets explored for the analysis are listed in Table \ref{M2.35 nuclear matter}. 
  The correlation matrix suggests that the symmetry energy ($J$) and its slope ($L$) exhibit relatively strong correlations with $M_{\mathrm{max}}$. At first glance, these parameters seem to play a major role. 
  Nonetheless, subsequent investigation indicates that their apparent correlations may be incidental rather than indicative of a true underlying influence on $M_{\mathrm{max}}$.

\begin{figure}[t]
    \centering
    \includegraphics[width=1.0\linewidth]{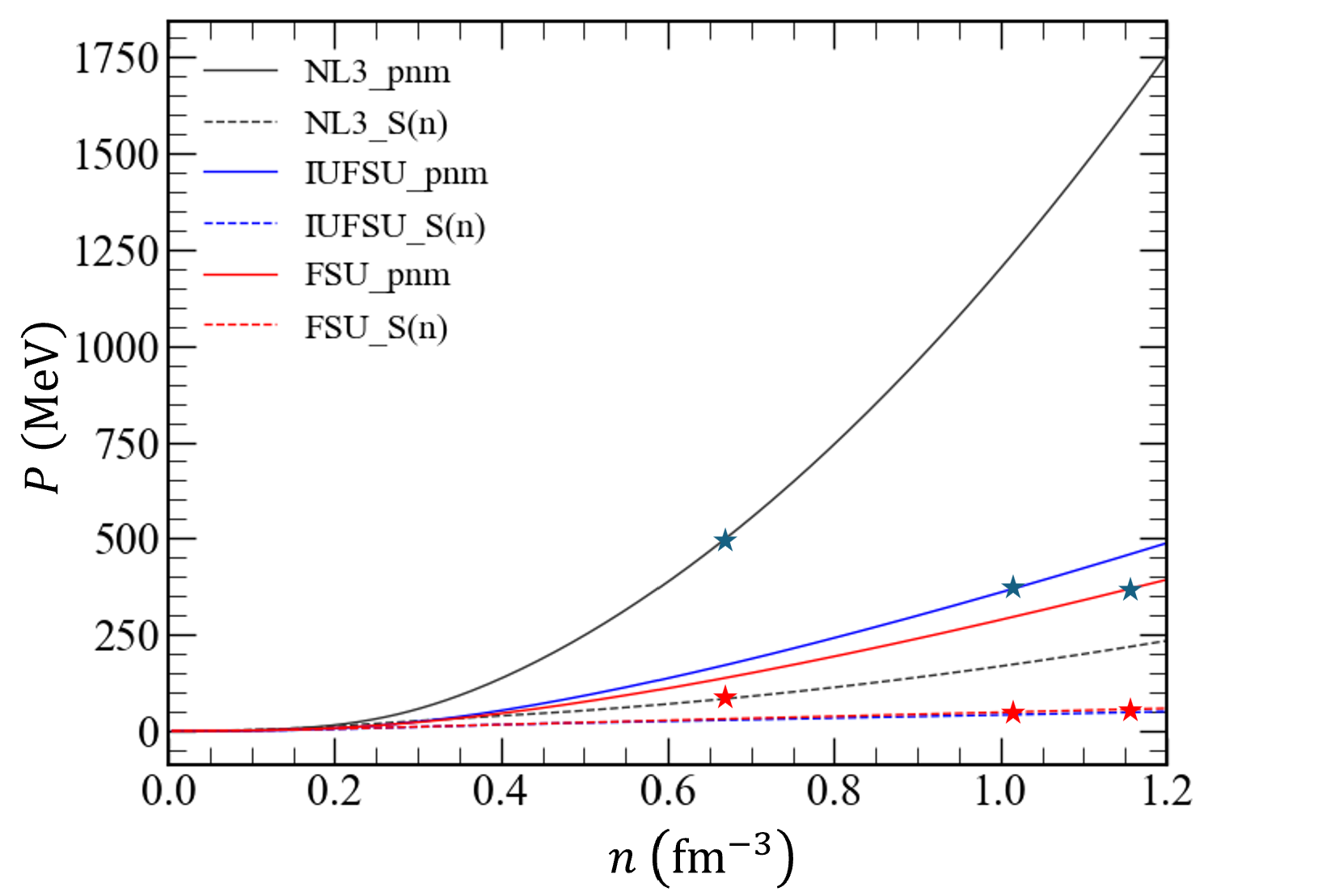}
    \caption{Pressure as a function of baryon density for the NL3, IUFSU, and FSU parameter sets. The total pressure of pure neutron matter and the symmetry energy contribution ($n^2\tfrac {S(n)}{dn}$)  are shown separately. The central densities of the corresponding maximum-mass neutron stars are indicated with stars.
}
    \label{fig:pressure by sym}
\end{figure}

  The equation of state of pure neutron matter is a good approximation to beta-equilibrium matter in a neutron star. Therefore, it is believed that understanding the density dependence of the symmetry energy is essential for reproducing an accurate mass–radius sequence of neutron stars. 
  This is true for predicting the radius of a typical-mass neutron star ($1.4\, M_\odot$), as can be seen from the correlation between the pressure of beta-equilibrium matter at saturation density and $R_{1.4}$ \cite{Lattimer2001,lattimer13,Lim2022}. In the vicinity of saturation density, the pressure of pure neutron matter can be approximated as follows:
\begin{equation}
\label{eq:pressure n0}
    P_{\rm pnm}(n) \simeq \frac{L}{3n_0}\,n^2+\frac{K+K_{sym}}{(3n_0)^2}(n-n_0)n^2,
\end{equation} 
where $K_{sym}= \big( 9n^2\frac{d^2S}{dn^2} \big)_{n=n_0} $. However, the maximum mass of a neutron star is governed by the pressure at high densities. 
In Figure \ref{fig:pressure by sym} we plotted the total pressure ($P_{pnm}(n)=P_{snm}(n)+n^2\tfrac{S(n)}{dn}$) as well as the contribution of $S(n)$ to the total pressure employing the widely used RMF parameter sets NL3, IUFSU, and FSU. The pressures corresponding to the central densities for each case are marked with a star. It is clearly observed that the contribution of $S(n)$ to the total pressure at high densities is relatively small. Therefore, it appears that the neutron star maximum mass is largely affected by the symmetric nuclear matter equation of state.

\begin{figure}[t]
    \centering
    \includegraphics[width=1.0\linewidth]{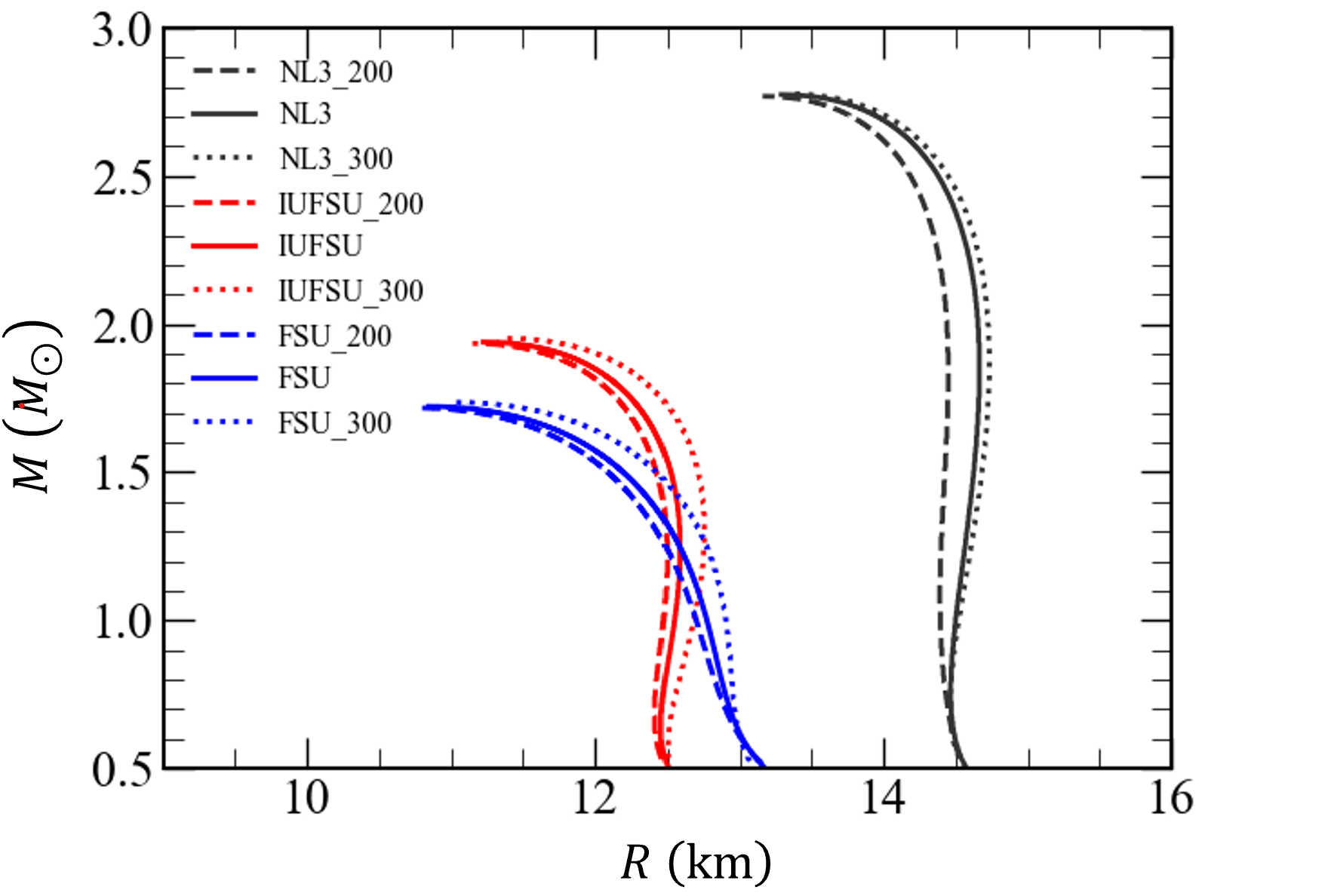}
    \caption{ Mass–radius relations for the NL3, IUFSU, and FSU parameter sets. For each case, additional curves are shown for varied incompressibility values of $K=200 \rm \, MeV$ and $K=300\,\rm MeV$ to illustrate the sensitivity of the stellar structure to the incompressibility of nuclear matter.}
    \label{fig:K dependence}
\end{figure}

Regarding the $\omega$ meson self coupling, for a given value of $\zeta$, which provides additional softening to the equation of state at high densities, the quartet of symmetric matter saturation properties - saturation density $n_0$, effective mass $m^*$, binding energy per nucleon $B/A$, and incompressibility $K$ can be used to algebraically determine the four parameters $\{g_s,g_v,\kappa,\lambda\}$.  
Within their empirically acceptable ranges, the impact of $B/A$ and $K$ on the maximum mass of neutron stars is found to be negligible. Obviously $B/A$ has nothing to do with the neutron star maximum mass since it does not generate any pressure ($ P =n^2 \tfrac{\partial (\epsilon/n)}{n} $). 
The influence of $K$ on the maximum mass of a neutron star appears to be important, since its contribution is not negligible even at high densities. To examine the impact of $K$ on the maximum mass of a neutron star, we plotted NL3, IUFSU, and FSU parameter sets with adjusted values of $K=200$  and $300$ MeV (Figure \ref{fig:K dependence}). Depending on the value of $K$, the radius of a typical-mass neutron star is notably affected. This phenomenon can be easily understood by referring to Eq.\ \eqref{eq:pressure n0}. Surprisingly, the neutron star maximum mass is rather insensitive even to unreasonably small (200 MeV) and large (300 MeV) incompressibility values. This indicates that the value of the incompressibility is not crucial for determining the maximum mass of a neutron star. 
According to the correlation analysis shown in Figure~\ref{fig:cor}, the skewness parameter ($Q$) also appears to have a strong correlation with the maximum mass ($M_{\mathrm{max}}$). However, $M_{\mathrm{max}}$ is also strongly correlated with the effective mass ($m^*$), which in turn shows a strong correlation with $Q$ in the same analysis. This suggests that the apparent dependence of $M_{\mathrm{max}}$ on $Q$ may primarily reflect its underlying correlation with $m^*$, implying that controlling $m^*$ may effectively amount to controlling $M_{\mathrm{max}}$.

With this observation, we fix
\begin{equation}\label{eq:incomp}
\frac{B}{A} = -16.3\ \mathrm{MeV}, \quad
K = 240\ \mathrm{MeV},    
\end{equation}
while varying
\begin{equation}
0.155 \le n_0 \le 0.165\ \mathrm{fm}^{-3}, \qquad
0.54 \le m^* \le 0.72.    
\end{equation}
At saturation ($n_3=0$),  $\frac{\epsilon}{n}=\frac{\partial \epsilon}{\partial n}$, which gives
    \begin{equation}
        M_N+B/A=V_0+\sqrt{k_f^2+M^{*2}}
    \end{equation}
    and along with Eq.\,\eqref{Eq euler Lagrange:omega}, $g_v$ is determined for given $\zeta$.
    From Eq.\,\eqref{Eq energy density}, we have
    \begin{equation}
        \begin{aligned}
(M_N+ B/A) n &= \tfrac{2}{\pi^2}\int_0^{k_f} k^2\sqrt{k^2+M^{*2}}dk \\
& + \tfrac{1}{2}\tfrac{m_s^2}{g_s^2}\Phi^2+\tfrac{\kappa}{3}\Phi^3+\tfrac{\lambda}{4}\Phi^4\\
& -\tfrac{1}{2}\tfrac{m_v^2}{g_v^2}V_0^2-\tfrac{\zeta}{4}V_0^2 +V_0 \, n .
\end{aligned} \label{Eq energy density:SNM}
    \end{equation}
From the definition of Eq.\,(\ref{Eq incompressibility}),
we have 
\begin{equation}
    \begin{aligned}
K &=  \frac{6k_f^3}{\pi^2} \frac{1}{(m_v^2/g_v^2)+3\zeta V_0^2 }  
 +\frac{3k_f^2}{\sqrt{k_f^2+M^{*2}}} \\
& -\frac{6k_f^3}{\pi^2}\frac{M^*}{\sqrt{k_f^2+M^{*2}}} \frac{g_s^2}{m_s^2}\\
 \times & \left[1+\frac{g_s^2}{m_s^2} \left( 2\kappa g_s\Phi+3\lambda g_s\Phi^2 +\frac{2}{\pi^2}\int_0^{k_f} \frac{k^4dk}{\sqrt{M^{*2}+k_f^2}}   \right) 
 \right] ^{-1}
    \end{aligned} \label{Eq incompressibility:formula}
\end{equation}
Then by combining Eqs.\,(\ref{Eq euler Lagrange:sigma}),(\ref{Eq effective mass}),(\ref{Eq energy density:SNM}) and (\ref{Eq incompressibility:formula}), the coupling constants $g_s,\kappa,\lambda$ are determined.

   \begin{figure}[t]  
  \centering
  \includegraphics[width=1.0 \linewidth]{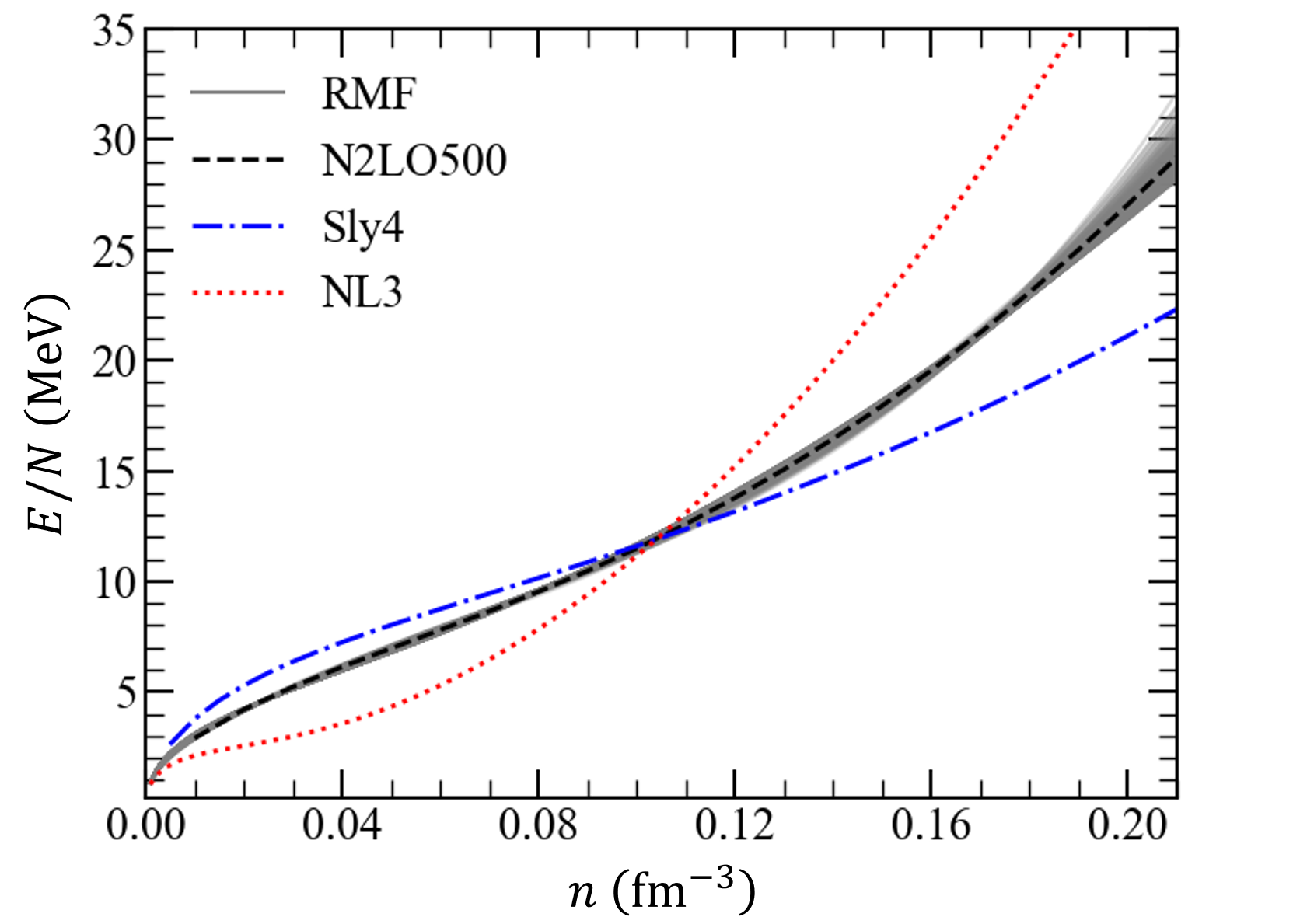}  
  \caption{
Energy per particle $E/N$ as a function of baryon number density $n$ for pure neutron matter. The gray band (labeled ``RMF'') represents a family of relativistic mean field (RMF) equations of state constrained to reproduce the microscopic results shown by the black dashed line (``N2LO500''). For comparison, results from two representative models-SLy4 (blue dot-dashed line) and NL3 (red dotted line)-are also shown.}
  \label{Fig PNM at low}
\end{figure} 

As for isospin-related parameters $g_\rho$ and $\Lambda_v$, rather than specifying the symmetry energy and the slope parameter, pure neutron matter (PNM) EOSs were constrained to the PNM EOS from chiral effective field theory (ChEFT) by Holt and Kaiser\, \cite{holt17prc} at low densities ($n\le0.21\,\rm fm ^{-3}$). To enforce the PNM constraint, we minimize the following chi-square objective function:
\begin{align*}
\chi^2 \; &=\;
\sum_{i=1}^{21}
\left(
 \frac{E_{\rm pnm}(n_i) - E_{\chi{\rm EFT}}(n_i)}{E_{\chi{\rm EFT}}(n_i)}
\right)^{\!2},
\end{align*}
where $E(n) = \frac{\varepsilon(n)}{n} - M_N$,
 $n_i = 0.01\,i\ \mathrm{fm}^{-3}$.

Figure \ref{Fig PNM at low} shows the energy per nucleon of pure neutron matter calculated using our RMF model parameters.
For comparison, we also include the PNM curves from NL3\,\cite{Lalazissis1997}, 
SLy4\,\cite{chabanat1997}, and chiral MBPT calculations with the N2LO 500 interaction\,\cite{holt17prc}.
Since our model is fitted to the chiral MBPT calculations, the figure demonstrates that our RMF model provides an excellent description of the chiral MBPT results.
Recent work by Reed et al.\ \cite{Reed2025ccn} applied RMF models to describe both finite nuclei and neutron matter using microscopic chiral EFT results, particularly the charge radii and neutron skin thicknesses of closed-shell nuclei.
This indicates that microscopic calculations can be well represented within the RMF formalism.
In contrast to their study, our aim is to focus on the nuclear matter properties and the maximum mass of neutron stars.

\section{Results}\label{sec:results}
Although massive hadrons such as hyperons might be energetically allowed in the core of neutron stars, their inclusion generally leads to a lower maximum mass for neutron stars compared to models without hyperons.
This has become particularly problematic in light of recent observations of massive neutron stars with masses around two solar masses, which challenge models that include such exotic particles. Therefore, a minimal model is often invoked, consisting of neutrons, protons, electrons, and muons in beta equilibrium. In this model, the fraction of each particle species relative to the total baryon density is determined by the following equations along with the charge neutrality condition.
\begin{align*}
\mu_n &= \mu_p + \mu_e, \\
\mu_e &= \mu_\mu.
\end{align*}
where $\mu_p$, $\mu_n$, $\mu_e$, and $\mu_\mu$ are the chemical potentials of the proton, neutron, electron, and muon, respectively. The proton and neutron chemical potentials are given by
\begin{align}
\mu_p &= \sqrt{k_p^2 + {M^*}^2}\;+\;V_0\;+\;\tfrac{B_0}{2},\\
\mu_n &= \sqrt{k_n^2 + {M^*}^2}\;+\;V_0\;-\;\tfrac{B_0}{2},
\end{align}
where $k_p$ and $k_n$ are fermi momenta for protons and neutrons, respectively.\\
On the other hand, the leptonic chemical potentials are given by non-interacting fermion, as follows:
\begin{equation}
\mu_{e,\mu} \;=\; \sqrt{m_{e,\mu}^2 + k_{e,\mu}^2}.
\end{equation}
By collecting contributions from nucleons and leptons, the total energy density and pressure are obtained:
\begin{align}
\epsilon &= \epsilon_{\rm nuc}(n, y_p) \;+\; \epsilon_e \;+\; \epsilon_\mu, \\
P        &= P_{\rm nuc}(n, y_p) \;+\; P_e \;+\; P_\mu,
\end{align}
where 
$y_p = \frac{n_p}{n}$, is the proton fraction of nuclear matter
and the leptonic contributions are
\begin{align}
\epsilon_{e,\mu}
&= \frac{1}{\pi^2}
   \int_0^{k_{e,\mu}}
   \sqrt{k^2 + m_{e,\mu}^2}\;k^2\,dk, 
   &\\
P_{e,\mu}
&= \frac{1}{3\pi^2}
   \int_0^{k_{e,\mu}}
   \frac{k^4}{\sqrt{k^2 + m_{e,\mu}^2}}\,dk,
\end{align}
with $k_e$, and $k_\mu$ denoting the Fermi momenta of electrons and muons, respectively. 

By solving the Tolman-Oppenheimer-Volkoff (TOV) equations given below 
for a given equation of state, the mass-radius relation of a static and spherically symmetric neutron star can be obtained.
\begin{equation}
\frac{dP}{dr} = -\frac{[P(r) + \epsilon(r)] \left[ M(r) + 4\pi r^3 P(r) \right]}{r \left[ r - 2M(r) \right]}\,,
\end{equation}
\begin{equation}
\frac{dM}{dr} = 4\pi r^2 \epsilon(r),
\end{equation}
where $M(r)$ is the enclosed mass for a radius $r$.
\begin{figure}[t]  
  \centering
  \includegraphics[width=1.0\linewidth]{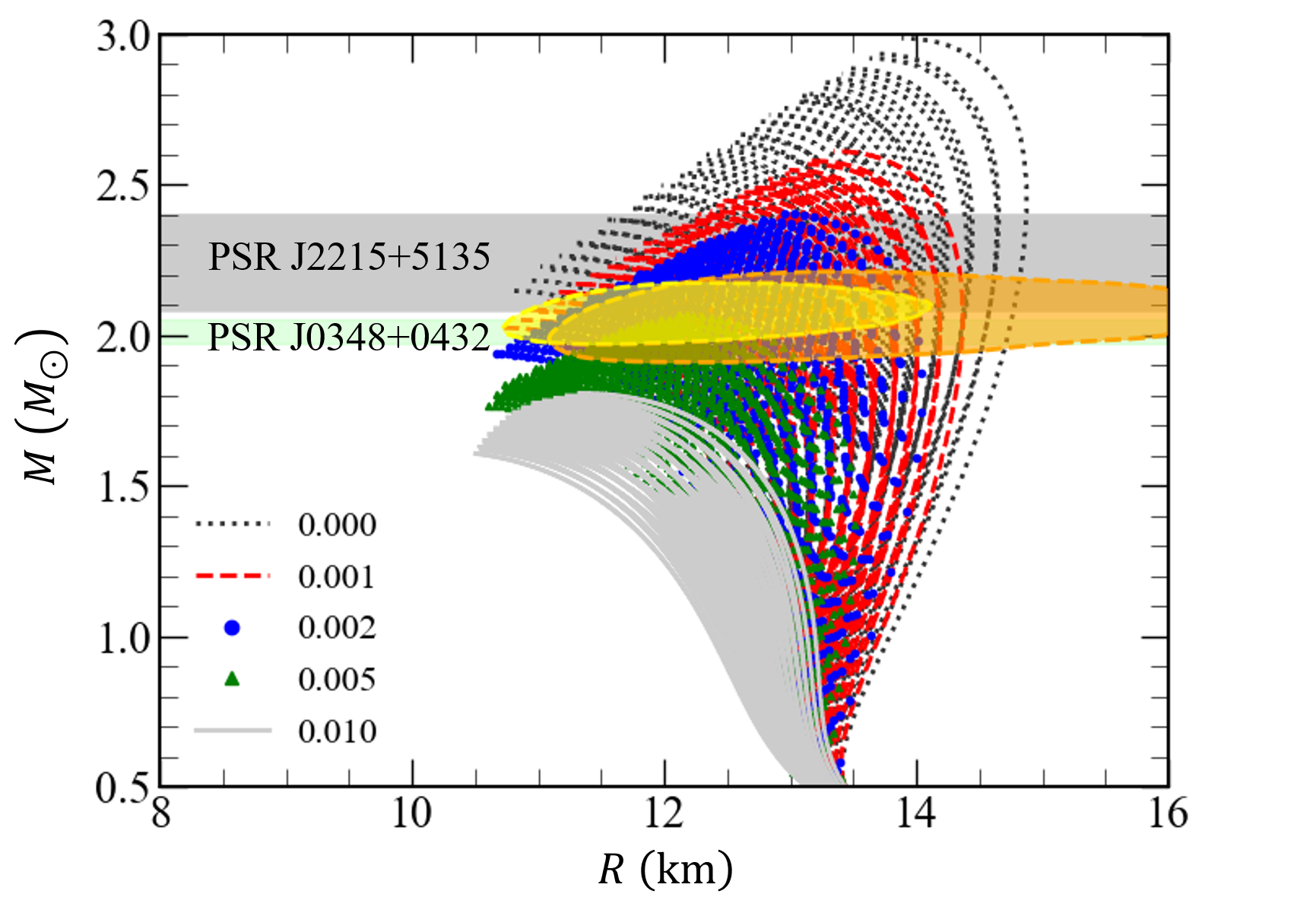}  
  \caption{Mass-radius relations for neutron stars with varying values of the coupling constant $\zeta$, as indicated in the legend. Each curve corresponds to a different equation of state model. Observational constraints from the massive pulsars PSR J2215+5135 and PSR J0348+0432 are shown as horizontal bands \cite{Linares2018,Antoniadis2013}. Results from the NICER analyses of PSR J0740+6620 by Riley et al. and Miller et al. are shown as dotted yellow and orange contours, respectively \cite{Riley2021,Miller2021}. The enclosed areas indicate the $+1\,\sigma$ confidence regions.} 
  \label{Fig MR all}
\end{figure}
Since the effect of the crust equation of state on the maximum mass of neutron stars is negligible, the BPS equation of state is adopted in this work \cite{BPS}.

Figure \ref{Fig MR all} shows the mass-radius relations for all sequences constrained by the chiral pure neutron matter equation of state at low densities. Sequences with the same magnitude of the vector meson self-coupling $\zeta$ are indicated by the same color. The self-coupling constant $\zeta$ is varied over the values $0$, $0.001$, $0.002$, $0.005$, $0.01$. For a fixed value of $\zeta$, each EOS corresponds to a different combination of saturation density $n_0$ and effective mass $m^*$. The saturation density $n_0$ ranges from $0.145$ to $0.165\,\mathrm{fm}^{-3}$ in steps of $0.005\,\mathrm{fm}^{-3}$, and the effective mass ranges from 0.54 to 0.72 in steps of 0.02. 
For each value of $\zeta$, the leftmost mass-radius sequence corresponds to $n_0 =0.165 \rm \ fm^{-3}$ and $m^*=0.72$, while the rightmost sequence to $n_0 =0.145 \rm \ fm^{-3}$ and $m^*=0.54$. 
The incompressibility ($K$) and binding energy per nucleon ($B/A$) are fixed at $240$\,MeV and $-16.3$\,MeV, respectively.
Note that $B/A$ and $K$ have no strong correlations 
with the maximum mass of neutron stars as seen in the table\,\ref{tab:formula}
although they can affect the value of radius of a neutron star with canonical mass. Since the aim of this study is to explore an empirical relation for the maximum mass of neutron stars, the cases with $\zeta =0.01$ are excluded because none of RMF parameter sets with $\zeta = 0.01$ are able to predict a $ \sim2.0 \, \msun$ neutron star. 

\begin{figure*}[htbp]  
  \centering
  \includegraphics[width=0.9 \linewidth]{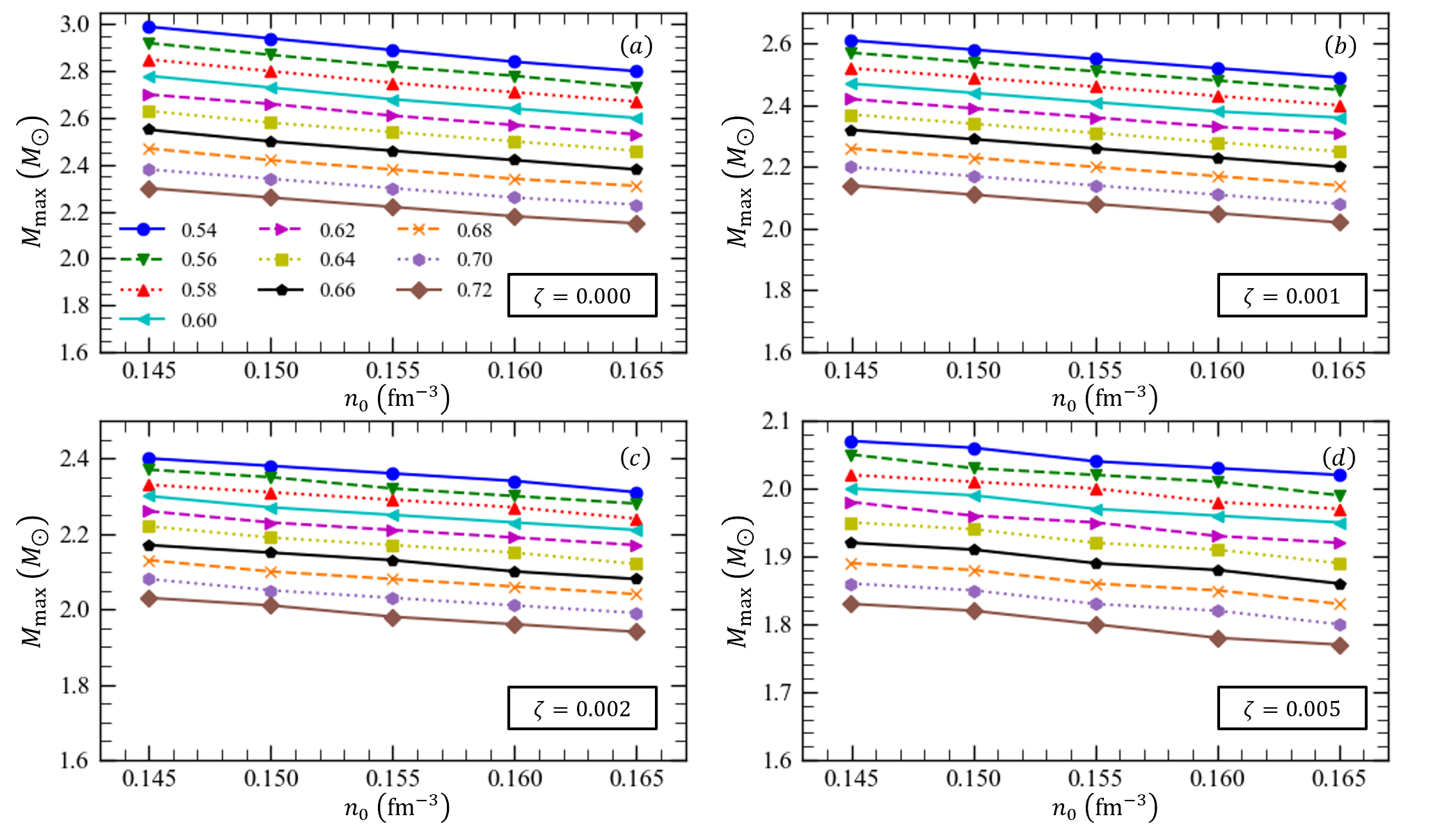}  
  \caption{
Maximum neutron star mass \(M_{\text{\rm max}}\) as a function of the nuclear saturation density \(n_0\) . In panels (a)–(d), different symbols represent results with the same effective mass \(m^*\). Each panel (a), (b), (c), and (d) corresponds to $\zeta=0$, $0.001$, $0.002$, $0.005$ respectively. 
}
  \label{Fig n0 Mmax}
\end{figure*}

\begin{figure*}[htbp]  
  \centering
  \includegraphics[width=0.9 \linewidth]{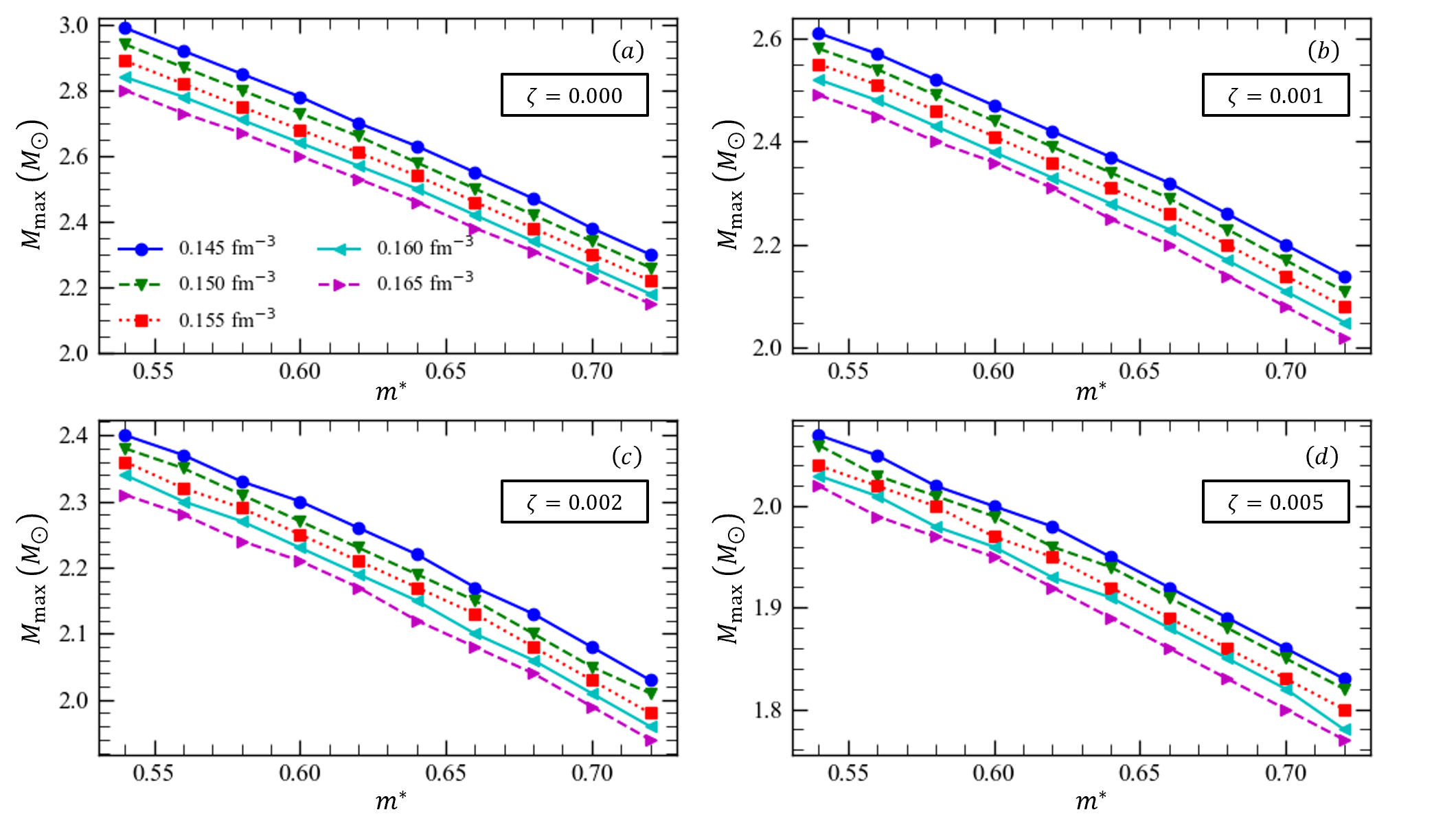}  
  \caption{
Maximum neutron star mass \(M_{\text{\rm max}}\) as a function of the effective mass \(m^*\). In panels (a)–(d), different symbols represent results with the same saturation density \(n_0\). Each panel (a),(b),(c), and (d) corresponds to $\zeta=0$, $0.001$, $0.002$, $0.005$ respectively. 
} 
  \label{Fig mstar Mmax}
\end{figure*}

Figures \ref{Fig n0 Mmax} and \ref{Fig mstar Mmax} show the maximum mass dependence on $n_0$  and $m^*$ while keeping $\zeta$ fixed. Note that there is a linear dependence on each parameter which suggests 
the empirical formula for the maximum mass of neutron stars as a function of $n_0$, $m^*$,
and $\zeta$. Figure \ref{Fig n0 Mmax} shows plots of $M_{\rm max}$ versus $n_0$, where instances with identical $m^*$  values are connected by each line. Within each panel, lines corresponding to smaller $m^*$ values predict progressively higher maximum masses. 
The sequence of lines from bottom to top corresponds to decreasing $m^*$ values from $0.72$ to $0.54$, with an interval of 0.02. The slopes of the lines within each panel are observed to be nearly uniform, suggesting that $\partial M_{\rm max}/\partial n_0 $ is not sensitive to $m^*$, but rather governed by $\zeta$. Thus, $\partial M_{\rm max}/\partial n_0 $ can be written as a function of $\zeta$, i.e., $\partial M_{\rm max}/\partial n_0 =f(\zeta)$.

Figure \ref{Fig mstar Mmax} presents plots of $M_{\rm max}$  versus $m^*$, where $n_0$ values range from 0.145 to 0.165 $\mathrm{fm}^{-3}$ in steps of 0.005 $\mathrm{fm}^{-3}$. Similar to the analysis in Figure \ref{Fig mstar Mmax}, each line connects instances with identical $n_0$ values. In each panel, the lines corresponding to lower $n_0$ values are positioned higher.
The slopes of these lines are observed to be almost constant, indicating that $\partial M_{\rm max}/\partial m^* $  is primarily influenced by $\zeta$ rather than $n_0$. Consequently, $\partial M_{\rm max}/\partial m^* $ can be expressed as a function of $\zeta$, i.e., $\partial M_{\rm max}/\partial m^*=g(\zeta) $. Taken into account these linear relations, 
the general functional form of $M_{\rm max}$ as a function of ($n_0$, $m^*$, $\zeta$) as can be inferred:
\begin{equation}
\label{Eq Mmax formula}
M_{\text{\rm max}} = f(\zeta)\, n_0 + g(\zeta)\, m^* + h(\zeta)\,.
\end{equation}

In order to investigate the dependence on $\zeta$, we adopted four sampling points: 0, 0.001, 0.002, 0.005. Based on this, the functions $f(\zeta)$, $g(\zeta)$, and 
$ h(\zeta)$ assumed to be quadratic in $\zeta$. With this form in hand, the specific functions were determined through a least squares fitting procedures. The determined coefficients for the functions $f(\zeta)$, $g(\zeta)$, and $h(\zeta)$ are listed below: 
\begin{equation}
\begin{aligned}
    \label{Eq Mmax formula fgh}
    f(\zeta) &= -123900\ \zeta^2 + 1084.4\ \zeta - 3.6812 \\
    g(\zeta) &= -312652\ \zeta^2 + 2681.2\ \zeta - 8.4824 \\
    h(\zeta) &= +148575\ \zeta^2 - 1335.3\ \zeta + 6.1959.
\end{aligned}
\end{equation}

To verify the effectiveness of this empirical maximum mass formula, we applied it to many existing RMF forces (see Table \ref{tab:formula}). 
Recently, studies have been carried out on the effects of the delta-meson term on neutron star properties, and it 
has been suggested that incorporating the delta meson into the RMF framework can enhance the maximum mass of a neutron star \cite{VirenderThakur2022,FanLi2022,Miyatsu2023}. 
\begin{figure}[t]  
  \centering
  \includegraphics[width=1.0\linewidth]{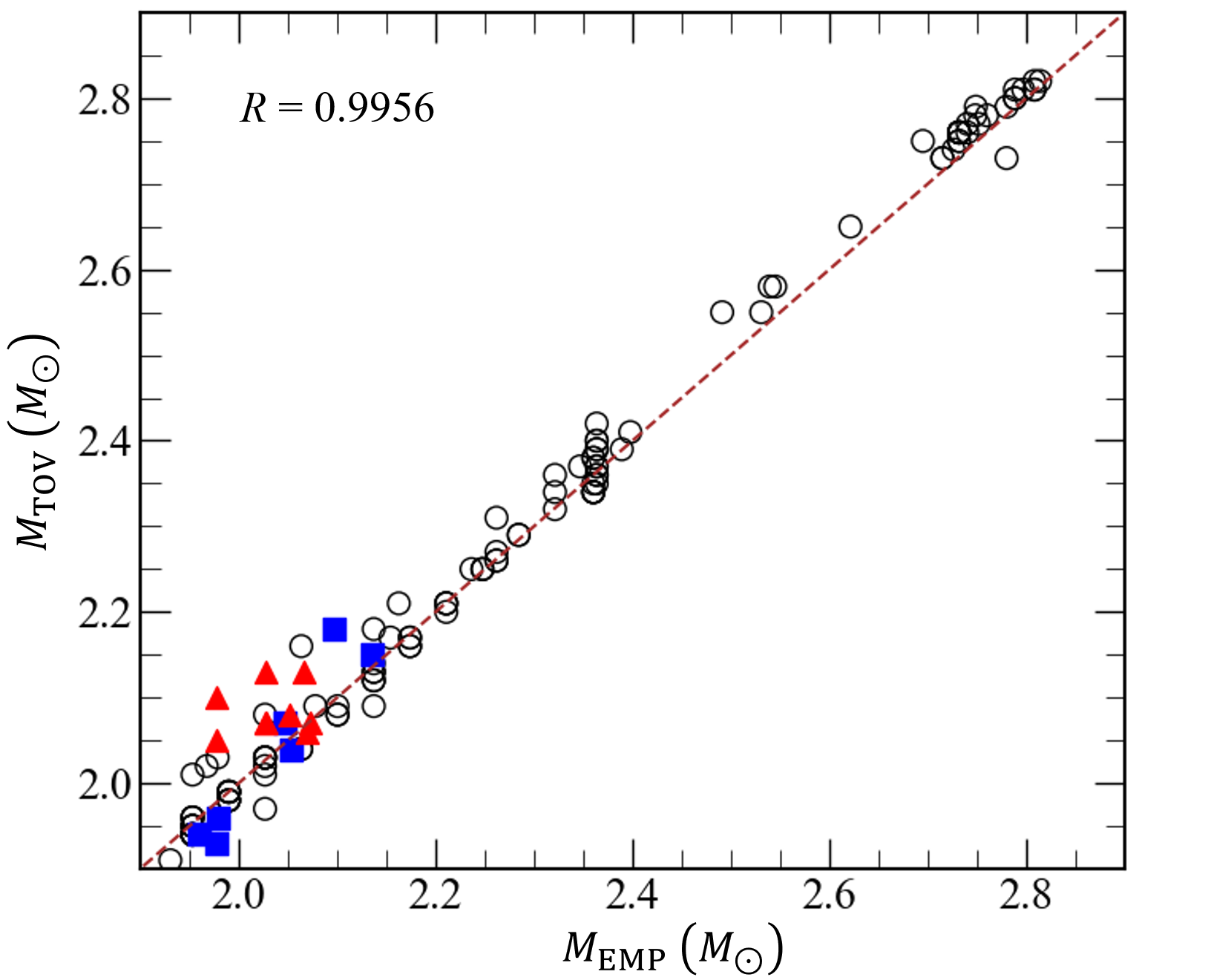}  
  \caption{
Comparison between the maximum neutron star mass $M_{\text{\rm max}}$ of existing RMF EOSs obtained by solving TOV equation $M_{\rm TOV}$ and the values predicted by the proposed empirical formula $M_{\rm EMP}$ (Table \ref{tab:formula}). 
Black circles ($\zeta=0$) and blue squares($\zeta \not = 0$) correspond to RMF models without the $\delta$ meson, while red triangles represent models including the $\delta$ meson. The Pearson correlation coefficient is shown in the upper left corner of the figure.
}
  \label{Fig formula}
\end{figure}
In particular, including the sigma–delta coupling was found to be very effective in enhancing the maximum mass while simultaneously reducing the radius of canonical neutron stars. As a result, the error associated with the empirical formula becomes larger compared to the case without the delta meson. This increase in error is likely due to the splitting of the effective mass induced by the delta meson, which results in a lower neutron effective mass. 

Table \ref{tab:formula} lists not only $n_0$, $m^*$  and $\zeta$  but also various other nuclear matter properties. However, it is found that, apart from these three quantities, the other properties have little impact on the maximum mass, indicating that $M_{\rm max}$ can be accurately predicted by the empirical formula. This implies that, within the RMF framework, the maximum mass of a neutron star can be easily estimated through the empirical formula without solving the TOV equations for each given Lagrangian density. 
To visually demonstrate the accuracy of our formula Eqs.(\ref{Eq Mmax formula}) and\,(\ref{Eq Mmax formula fgh}), we plotted the maximum mass $M_{\rm TOV}$ obtained by solving the TOV equations against the corresponding $M_{\rm RMF}$ estimated from the formula (see Figure \ref{Fig formula}). The data points align closely along the one-to-one line with a Pearson correlation coefficient of 0.9956, illustrating the strong agreement between them.
As $\zeta$ becomes too large, the prediction from the formula become inaccurate. For example, the FSU model has a large value of $\zeta=0.01$, and its predicted $M_{\rm max}$ deviates significantly from the formula. Although it is possible to expand the fitting range of $\zeta$ to include such models, doing so would increase the overall fitting error. Moreover, there is no compelling reason to include models with unexpectedly small $M_{\rm max}$. Therefore, this formula is best applied to models with $\zeta \lesssim 0.005 $, 
which is the critical values for achieving $M_{\rm max} > 2.0\,\msun$.

Within the RMF framework for homogeneous nuclear matter, the $\sigma$ meson sector is characterized by three parameters $(g_s,\kappa,\lambda)$, while the $\omega$ meson sector involves two parameters $(g_v,\zeta)$. Although the $\rho$ meson terms contribute to the symmetry energy, their impact on equation of state at high densities is relatively minor compared to that of the sigma and omega mesons. Consequently, the $\sigma$ and $\omega$ mesons predominantly determine the properties of the system at high densities and even the contributions from $\rho-\sigma$ mixing and $\rho-\omega$ mixing terms are negligibly small due to the small value of $B_0$ compared to $\Phi$ and $V_0$ at high densities. 
Thus, the remaining five degrees of freedom $(g_s,\kappa,\lambda,g_v,\zeta)$ can be associated with the determination of five nuclear matter properties. By treating $\zeta $ separately, the remaining four degrees of freedom are equivalent to fix the incompressibility $K$, the saturation density $n_0$, the effective mass $m^*$, and the binding energy per nucleon $B/A$. Among these, $K$ and $B/A$ have limited influence on the high-density behavior, leaving $n_0$ and $m^*$ as the two critical nuclear matter properties. As a result, $\zeta$ remains as an independent parameter controlling the high-density stiffness of the equation of state.

\begin{figure}[htbp]  
  \centering
  \includegraphics[width=0.8 \linewidth]{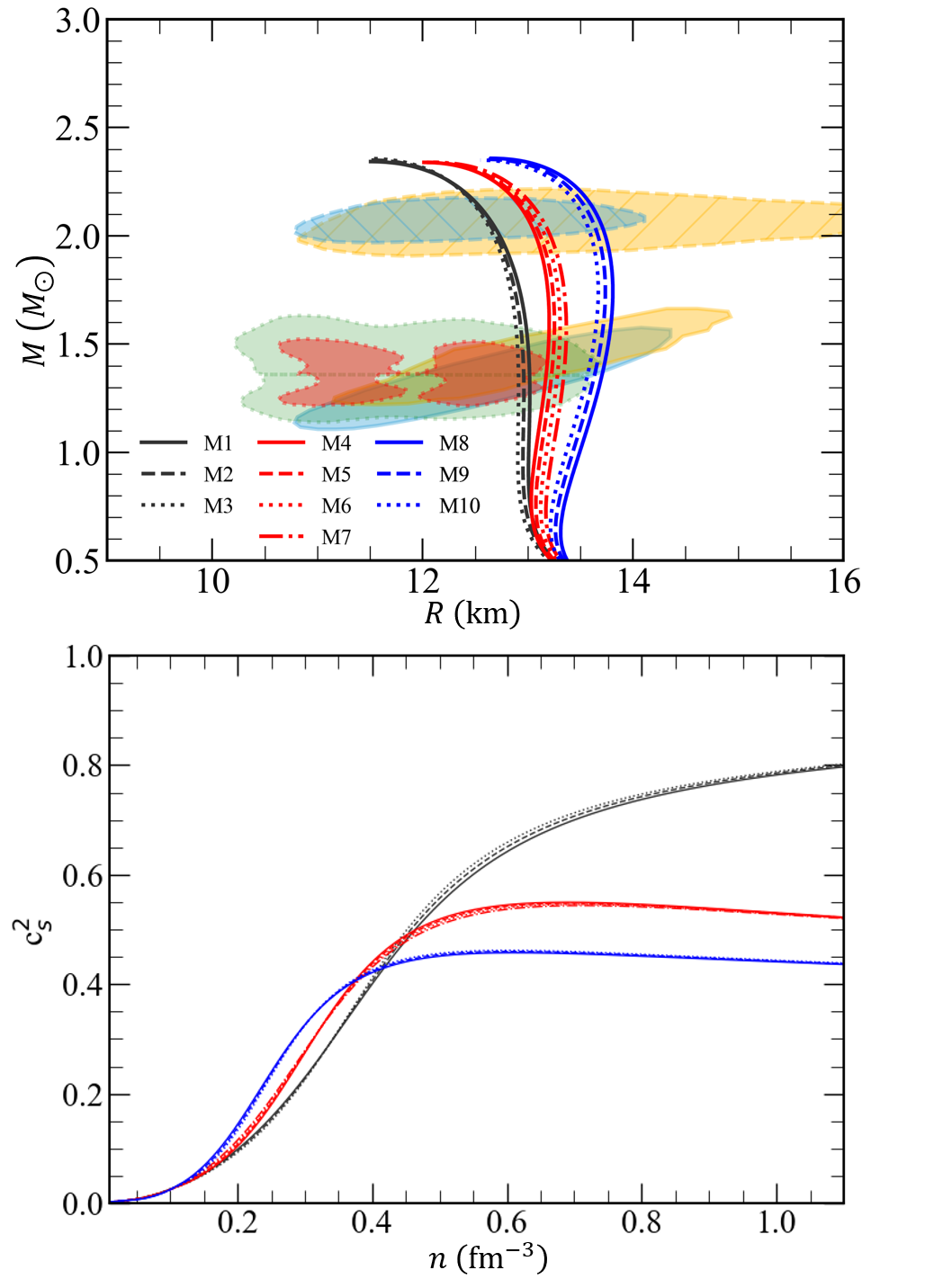}  
  \caption{(Upper) Mass–radius relations for neutron stars calculated from ten different EOS models (M1–M10). The models are grouped according to the coupling strength $\zeta$: M1–M3 (black) for $\zeta = 0$, M4–M7 (red) for $\zeta = 0.001$, and M8–M10 (blue) for $\zeta = 0.002$.
  The green and red shaded regions represent the 50\% and 90\% credible intervals from the gravitational-wave event GW170817 \cite{abbott2018b}. The yellow and blue hatched regions indicate the $1\,\sigma$ credible intervals from the NICER analyses of PSR J0740+6620 by Miller et al. and Riley et al. \cite{Miller2021,Riley2021}. Similarly, the yellow and blue-shaded represent the $1\,\sigma$ constraints from PSR J0030+0451 obtained by Miller et al. and Riley et al. respectively \cite{Miller2019,Riley2019}.
(Lower) The squared speed of sound $c_s^2$ as a function of baryon number density $n$ for the same models.
}
  \label{Fig M2.35}
\end{figure}

\begin{figure}[h]
  \centering
  \includegraphics[width=0.8\linewidth]{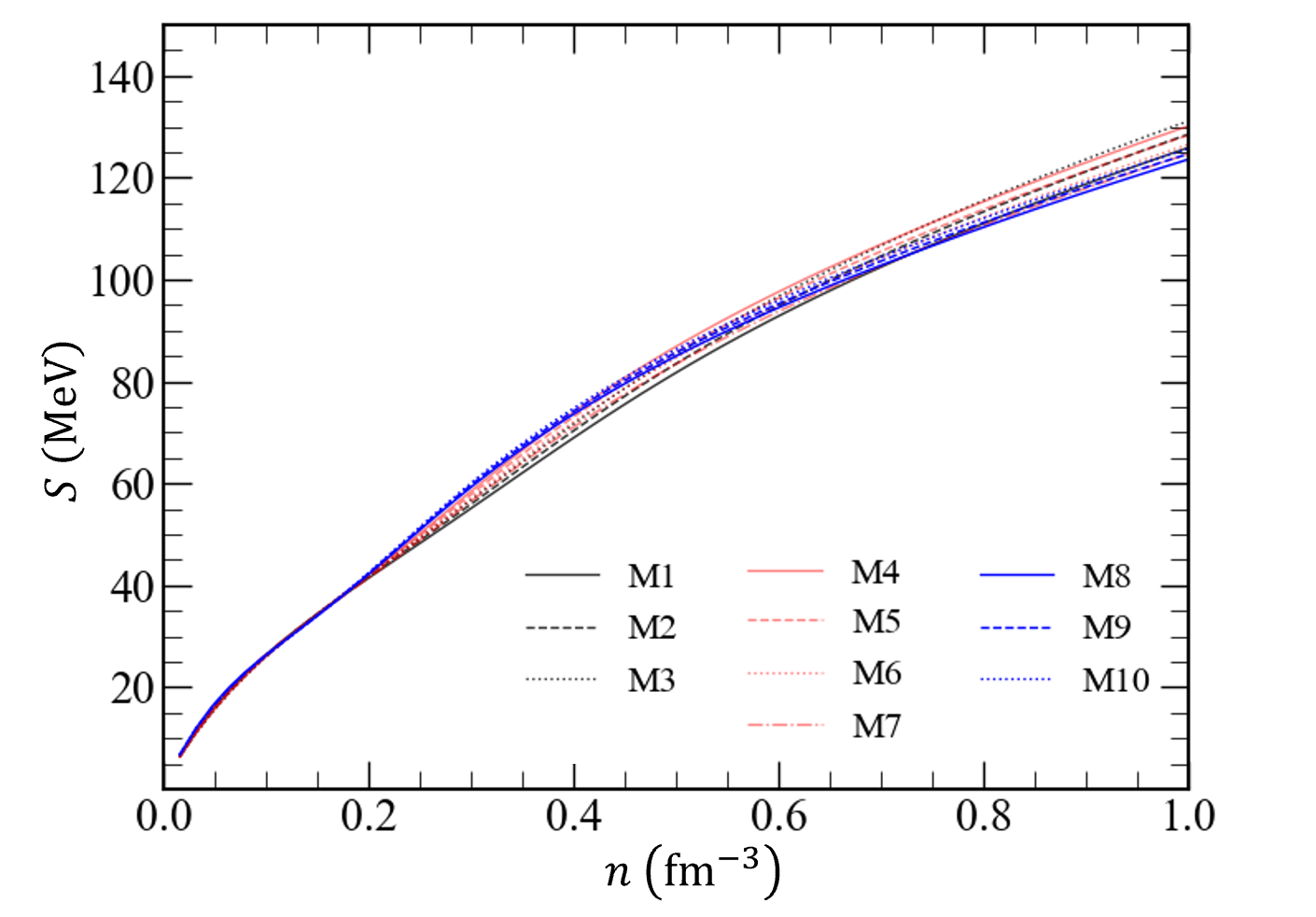}  
  \caption{Symmetry energy $S(n)$ as a function of baryon number density $n$ for ten theoretical EOS models (M1--M10), each consistent with a maximum neutron star mass of approximately $2.35\,M_\odot$. The models are grouped as M1--M3 (black), M4--M7 (red), and M8--M10 (blue), and are distinguished by different line styles. }
  \label{Fig M2.35_Sym}
\end{figure}

Our empirical maximum mass formula allows us to assess the capability of the RMF framework to predict highly massive neutron stars. As an example, we consider the case of $M_{\rm max} \sim 2.35\ M_\odot$.
Figure\,\ref{Fig M2.35} presents the mass–radius relations for each EOS model, showing that the maximum mass, $M_{\rm max}$, is approximately $2.35\,M_\odot$. The corresponding sound speed as a function of density in $\beta$-equilibrated matter is also displayed. Figure\,\ref{Fig M2.35_Sym} illustrates the density dependence of the symmetry energy for the same set of models.

In the analysis, three values of $\zeta$ were chosen: 0, 0.001, 0.002. For each value of $\zeta$, three cases of $(n_0,m^*)$ were selected for $\zeta=0$, $0.002$ and four cases for $\zeta=0.001$, and the results of the TOV calculations for these cases are listed in Table \ref{M2.35 nuclear matter}. The corresponding RMF parameters are shown in Table \ref{M2.35 nuclear matter}. The EOSs were similarly constrained by the chiral EFT PNM EOS with the binding energy per nucleon fixed at $B/A=-16.3\,\mathrm{MeV}$ and the incompressibility set to $K=240\,\rm{MeV}$. Since the saturation density values of many successful RMF forces are around $0.150\ \mathrm{fm}^{-3}$, only values between $0.145\ \mathrm{fm}^{-3}$ and $0.160\ \mathrm{fm}^{-3}$ are considered.

Although the symmetry energy and the slope parameters appear to have different values, this is merely a consequence of the difference in $n_0$. In fact, the density dependence of the symmetry energy is very similar across the different cases, as can be seen in Figure \ref{Fig M2.35_Sym}, where we plot the density-dependent symmetry energy for the ten models M1-M10. This is rather expected, since all cases share the same PNM EOS up to approximately $n_0 \sim 0.21\,\mathrm{fm}^{-3}$. EOSs with the same $\zeta$ exhibit a high degree of similarity. For cases with identical $\zeta$, the maximum mass $M_{\rm max}$ and the corresponding radius $R_{\rm max}$ are nearly the same, while differences mainly appear in $R_{1.4}$. As $\zeta$ increases, the increase in $R_{\rm max}$ becomes more pronounced, accompanied by a corresponding increase in $R_{1.4}$. This behavior can be understood from the $\frac{dP}{d\epsilon}$ versus baryon density plot shown in the lower panel of Figure \ref{Fig M2.35}, 
where the magnitude of the sound speed at low densities is larger with increasing $\zeta$, while it becomes smaller at higher densities. 
Among the considered models (M1-M10), all parameter sets successfully reproduce the mass-radius constraints inferred from the NICER analyses of PSR J0740+6620 and PSR J0030+0451 within the $1\, \sigma$ credible intervals. Meanwhile, the parameter sets M8-M10 are excluded by the GW170817 $90\,\%$ credible region and the parameter sets M4-M7 fail to lie within the $50\,\%$ credible region. 
Compared to numerous studies that found a strong correlation between the symmetry energy slope parameter ($L$) and $R_{1.4}$ \cite{alam16,lim2024,lim2025a},
the above analysis shows that for cases with similar symmetry energy density dependence, $R_{1.4}$ can be effectively tuned by varying $\zeta$ while keeping $M_{\rm max}$ fixed.
\setlength{\tabcolsep}{12pt}
\begin{table*}[htbp]
\centering
\caption{Nuclear matter properties and vector meson self-coupling ($\zeta$) for the model M1-M10. $M_{\rm max}$ is the maximum mass (in $M_\odot$), $R_{\rm max}$ and $R_{1.4}$ are radii (in km).\label{M2.35 nuclear matter}}
\begin{tabular}{lccccccccc}
\hline
\hline
Model & $\zeta$ & $n_0$ & $m^*$ & $J$ (MeV) & $L$ (MeV) & $M_{\text{\rm max}}$ & $R_{\text{\rm max}}$ (km) & $R_{1.4}$ (km) \\
\hline
M1  & 0     & 0.14961 & 0.70 & 33.4  & 63.05 & 2.343 & 11.51 & 13.01 \\
M2  & 0     & 0.15395 & 0.69 & 34.02 & 66.24 & 2.350 & 11.50 & 12.96 \\
M3  & 0     & 0.15829 & 0.68 & 34.64 & 69.51 & 2.357 & 11.49 & 12.91 \\
M4  & 0.001 & 0.15904 & 0.62 & 34.53 & 69.94 & 2.340 & 12.01 & 13.18 \\
M5  & 0.001 & 0.15459 & 0.63 & 33.877& 66.46 & 2.339 & 12.04 & 13.23 \\
M6  & 0.001 & 0.15014 & 0.64 & 33.231& 63.05 & 2.339 & 12.07 & 13.29 \\
M7  & 0.001 & 0.14569 & 0.65 & 32.598& 59.69 & 2.339 & 12.10 & 13.35 \\
M8  & 0.002 & 0.14760 & 0.56 & 32.412& 62.71 & 2.358 & 12.65 & 13.73 \\
M9  & 0.002 & 0.15220 & 0.55 & 33.106& 66.35 & 2.354 & 12.60 & 13.66 \\
M10 & 0.002 & 0.15679 & 0.54 & 33.808& 69.99 & 2.350 & 12.55 & 13.59 \\
\hline
\end{tabular}
\end{table*}
\setlength{\tabcolsep}{12pt}
\begin{table*}[htbp]
\centering
\caption{ Coupling constants for the models M1–M10. The table lists the scalar coupling ($g_s$), scalar self-interaction parameters ($\kappa$, $\lambda$), vector coupling ($g_v$), isovector-vector coupling ($g_\rho$), mixed coupling constants ($\Lambda_{s1}$, $\Lambda_{s2}$), and the vector self-interaction parameter ($\zeta$).}
\label{M2.35 parameters}
\begin{tabular}{lccccccccc}
\hline
\hline
Model & $g_s$ & $\kappa\ (\mathrm{fm}^{-1})$ & $\lambda$ & $g_v$ & $g_\rho$ & $\Lambda_{s1}$ & $\Lambda_{s2} $ & $\zeta$ \\
\hline
M1  & 9.1375 & 16.0803 & -31.3997 & 10.7474 & 12.1232 & 0.01897 & 0.02535 & 0     \\
M2  & 9.1236 & 15.4343 & -31.6206 & 10.7843 & 11.9561 & 0.02038 & 0.02172 & 0     \\
M3  & 9.1104 & 14.8396 & -31.6945 & 10.8183 & 11.7906 & 0.02157 & 0.01855 & 0     \\
M4  & 9.7500 & 11.8063 & -26.8022 & 12.0439 & 11.7345 & 0.01559 & 0.02205 & 0.001 \\
M5  & 9.7745 & 12.0655 & -26.4619 & 12.0462 & 11.8516 & 0.01389 & 0.02473 & 0.001 \\
M6  & 9.8012 & 12.3702 & -26.1283 & 12.0482 & 11.9758 & 0.01208 & 0.02768 & 0.001 \\
M7  & 9.8302 & 12.7220 & -25.7845 & 12.0500 & 12.1057 & 0.01013 & 0.03093 & 0.001 \\
M8  & 10.9508 & 10.8691 & -21.6550 & 13.9888 & 12.3372 & 0.00228 & 0.03592 & 0.002 \\
M9  & 10.9175 & 11.1060 & -23.4670 & 13.9403 & 12.3854 & 0.00536 & 0.03309 & 0.002 \\
M10 & 10.8893 & 11.3933 & -25.3240 & 13.8944 & 12.4504 & 0.00828 & 0.03053 & 0.002 \\
\hline
\end{tabular}
\end{table*}

\begin{figure*}[htbp]
\centering
\includegraphics[scale=0.25]{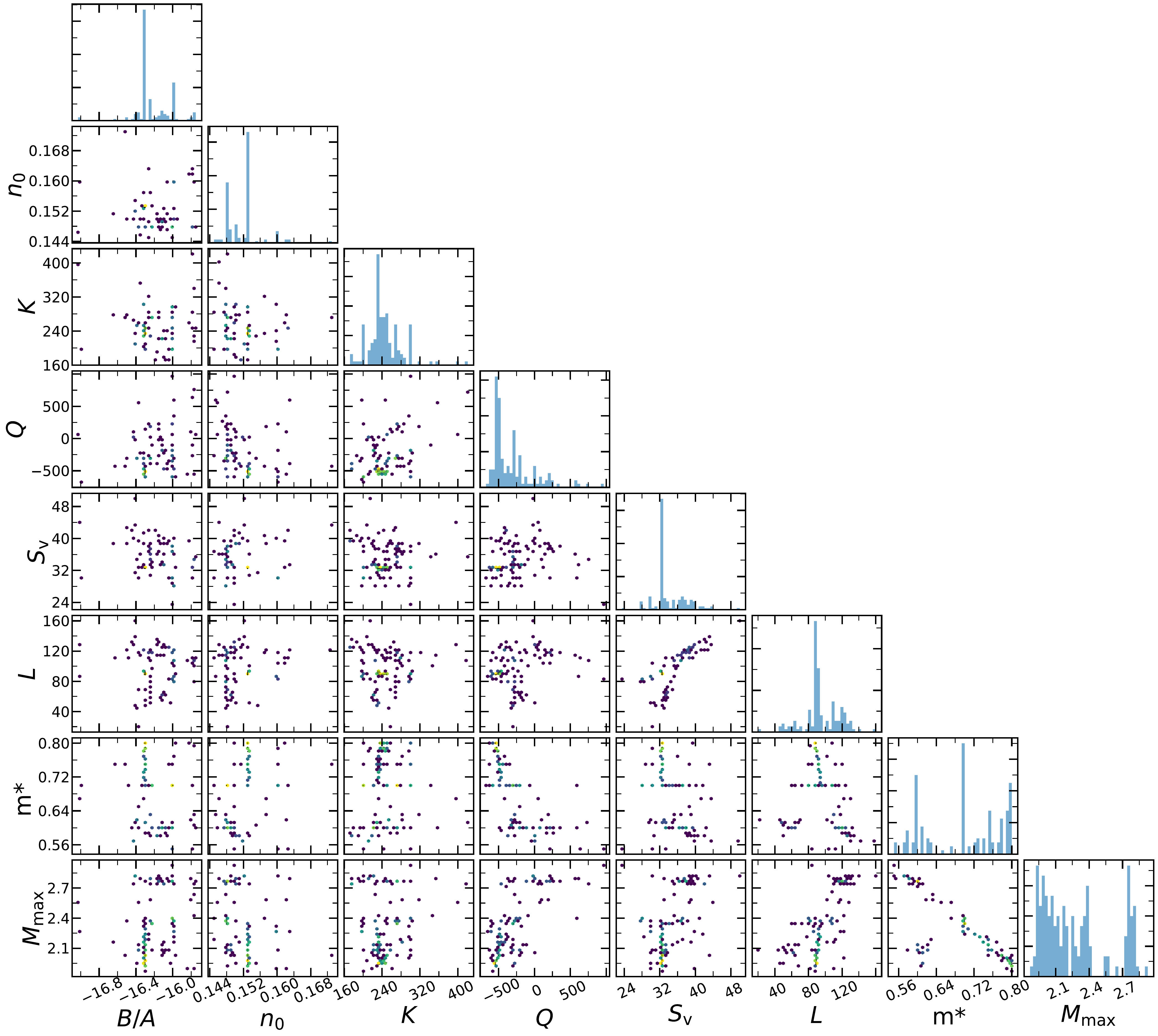}    
  \caption{Scatter plots of various nuclear matter properties obtained from 148 RMF models. The saturation density $n_0$ is given in fm$^{-3}$. The binding energy per nucleon $B/A$, incompressibility $K$, skewness $Q$, symmetry energy $S_v$, and slope parameter $L$ are given in MeV. The maximum mass $M_{\rm max}$ is given in solar masses ($M_\odot$)
  }
  \label{fig:rmfscatter}
\end{figure*}

As can be seen in Figure \ref{fig:rmfscatter}, there is a strong anti-correlation between $m^*$ and $M_{\rm max}$, even though the distribution of the maximum mass alone does not show any clear trend.
This also suggests the existence of an empirical relation between the maximum mass of neutron stars and the effective mass.
Note that the effective mass used in this figure is the Dirac effective mass.
In the RMF models, there are almost no correlations among nuclear matter properties, except between $S_v$ and $L$.
These variables, however, do not exhibit any correlation with the maximum mass of neutron stars.

\begin{figure*}[htbp]
\centering
\includegraphics[scale=0.25]{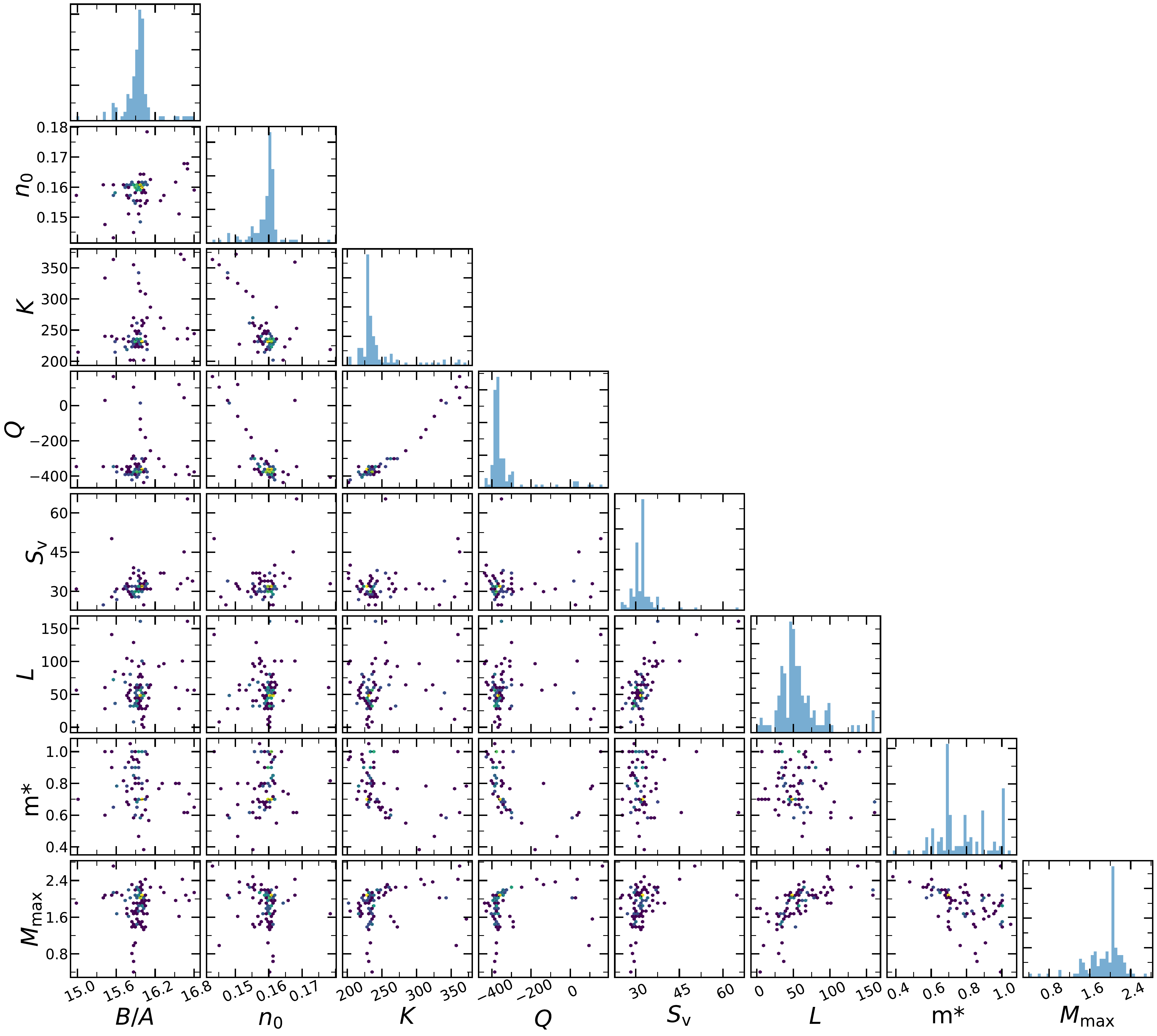}
\caption{The same plots as in Fig.,\ref{fig:rmfscatter} using 115 Skyrme force models.
}
\label{fig:skyrmescatter}
\end{figure*}

In the case of Skyrme-type nonrelativistic models, we do not find any empirical relation between the maximum mass of neutron stars and combinations of nuclear matter properties.
For example, Figure \ref{fig:skyrmescatter} presents the same scatter plots as in Figure \ref{fig:rmfscatter}, except now for Skyrme interactions.
It shows that $B$, $n_0$, and $K$ lie around well-averaged values \cite{dutra12,lim19e}.
In the Skyrme models, the effective mass corresponds to the Landau effective mass rather than the Dirac effective mass, with $m^*_L = \sqrt{k_F^2 + m_D^{*2}}$.
Even after converting to the Dirac effective mass, no correlation is obtained.


\section{conclusion}
\label{sec:conclusion}

In response to recent observational data, new RMF forces have been proposed to cope with the high maximum mass of neutron stars. Nevertheless, a detailed analysis of which factors within the RMF framework primarily determine the maximum mass has not yet been carried out. The
density dependence of the symmetry energy is believed to be essential ingredient for understanding the properties of neutron stars. Indeed, it is crucial for determining the radius of a typical-mass neutron star. However, with regard to the maximum mass of a neutron star, we have shown that the impact of the density dependence of the symmetry energy is rather insignificant. This suggests that the equation of state of symmetric nuclear matter may be highly relevant for understanding the maximum mass of a neutron star.

In the present work, we explored RMF models with five relevant parameters: $g_s$, $\kappa$, $\lambda$, $g_v$, $\rm$, and $\zeta$ from which the properties of symmetric nuclear matter can be algebraically determined. Since no pressure is deduced by the binding energy per baryon, $B/A$ at the saturation, it was excluded in deriving the universal formula. Even the influence of the incompressibility $K$ turned out to be insignificant for determining the neutron star maximum mass. By systematically generating 250 RMF EOSs with different combinations of $n_0$, $m^*$,  and $\zeta$, and calculating the corresponding maximum masses through the TOV equations, we derived an empirical formula for the neutron star maximum mass valid within the RMF framework. Its effectiveness was further demonstrated by applying it to a wide range of existing RMF forces, where it was found to reproduce the maximum mass values with good accuracy. Interesting, the maximum mass was found to be well predicted by the saturation density $n_0$, the effective mass $m^*$, and the omega meson self-coupling $\zeta$. In contrast, the neutron star maximum mass is less sensitive to the density dependence of the symmetry energy, the binding energy per nucleon $B/A$, and the incompressibility $K$ as supported by a variety of existing RMF parameter sets.

The correlation between $R_{1.4}$ and the density dependence of the symmetry energy has been significantly studied along with the neutron skin thickness, and it was suggested the accurate observation of typical neutron stars could constrain $L$ \,\cite{Lattimer2001,Lattimer2007, Horowitz2001b,Reed2021,FanJi2019}. However, $R_{1.4}$ can vary significantly depending on the value of $\zeta$, suggesting that this dependence should also be taken into account when constraining $L$ based on observational data. Indeed, an analysis based on the empirical formula for the case of $M \sim 2.35 M_\odot $ reveals that, when the maximum mass is fixed, both $R_{\rm max}$ and $R_{1.4}$ systematically increase with increasing $\zeta$, supporting the need to consider this effect. 
A systematic investigation to find the universal relation for the maximum mass of NS  established here with predictions for finite nuclei properties would also be a possible direction for future research.
\\
\acknowledgments
G.~Nam and Y.~Lim were supported supported by the National
Research Foundation of Korea(NRF) grant funded by the
Korea government(MSIT)(No. RS-2024-00457037) and
by Global - Learning \& Academic research institution for Master's·PhD students, 
and Postdocs(LAMP) Program of the National Research Foundation of Korea(NRF) grant funded by the Ministry of Education(No.  RS-2024-00442483).
Y. Lim was also supported by the Yonsei University Research Fund of 2025-22-0040.  Master project. J.~W. Holt was supported in part by the National Science Foundation under grant No.\ 2514930.

\setlength{\tabcolsep}{6pt}
\renewcommand{\arraystretch}{1.4}
\LTcapwidth=\textwidth

\begin{longtable*}{lcccccccccccc}

\caption{ Comparison of the maximum mass formula ($M_{\rm EMP}$) with TOV solution ($M_{\rm TOV}$) for existing RMF models. Nuclear saturation properties and $c_\omega$
are included to examine how the maximum mass are related with them. 
 $J$, $L$, $K$, $Q$, and $B/A$ are given in $\rm{MeV}$. Saturation density $n_0$ is given in $\rm{fm}^{-3}$. Formula and $M_{\rm{max}}$ are given in $M_\odot$. Error is given by $\frac{M_{\rm EMP}-M_{\rm TOV} }{M_{\rm TOV}}\times 100$.}
\label{tab:formula}
\\
\hline
 Models & $J$ & $L$  & $K$  & $Q$  & $B/A$  & $n_0 $ & $m^*$ & $\zeta$ & $M_{\rm EMP}$  & $M_{\text{\rm TOV}}$ & Error (\%) & Ref. \\
\hline
\endfirsthead
\multicolumn{13}{c}{TABLE III  $Continued.$}
\\
\hline
Models & $J$ & $L$  & $K$  & $Q$  & $B/A$  & $n_0 $ & $m^*$ & $\zeta$ & $M_{\rm EMP}$  & $M_{\text{\rm max}}$ & Error (\%) & Ref. \\
\hline
\endhead

\hline
\endfoot

\hline
\endlastfoot

CS	&	40.91 	&	131.75 	&	187.21 	&	-268.56 	&	-16.17 	&	0.150 	&	0.58 	&	0	&	2.788 	&	2.802 	&	0.47 &\,\cite{Rufa1988} 	\\
E	&	38.58 	&	124.42 	&	221.43 	&	9.74 	&	-16.13 	&	0.150 	&	0.58 	&	0	&	2.788 	&	2.806 	&	0.62 &\,\cite{Rufa1988} 	\\
ER	&	39.42 	&	126.95 	&	220.49 	&	-0.51 	&	-16.16 	&	0.149 	&	0.58 	&	0	&	2.797 	&	2.814 	&	0.62 &\,\cite{Rufa1988} 	\\
FAMA1	&	38.01 	&	120.50 	&	200.05 	&	-303.15 	&	-16.00 	&	0.148 	&	0.60 	&	0	&	2.732 	&	2.753 	&	0.77 &\,\cite{Piekarewicz2002} 	\\
FAMA2	&	38.01 	&	120.26 	&	225.07 	&	-117.68 	&	-16.00 	&	0.148 	&	0.60 	&	0	&	2.732 	&	2.755 	&	0.84 &\,\cite{Piekarewicz2002} 	\\
FAMA3	&	38.01 	&	120.03 	&	250.08 	&	52.64 	&	-16.00 	&	0.148 	&	0.60 	&	0	&	2.732 	&	2.757 	&	0.91 &\,\cite{Piekarewicz2002} 	\\
FAMA4	&	38.01 	&	119.79 	&	275.09 	&	208.07 	&	-16.00 	&	0.148 	&	0.60 	&	0	&	2.732 	&	2.760 	&	1.00 &\,\cite{Piekarewicz2002} 	\\
FAMA5	&	38.01 	&	119.55 	&	300.00 	&	348.76 	&	-16.00 	&	0.148 	&	0.60 	&	0	&	2.732 	&	2.762 	&	1.10 &\,\cite{Piekarewicz2002} 	\\
FAMB1	&	37.01 	&	108.40 	&	200.03 	&	-593.39 	&	-16.00 	&	0.148 	&	0.70 	&	0	&	2.364 	&	2.352 	&	-0.48 &\,\cite{Piekarewicz2002} 	\\
FAMB2	&	37.01 	&	108.21 	&	225.05 	&	-465.00 	&	-16.00 	&	0.148 	&	0.70 	&	0	&	2.364 	&	2.362 	&	-0.06 &\,\cite{Piekarewicz2002} 	\\
FAMB3	&	37.01 	&	108.03 	&	250.05 	&	-355.30 	&	-16.00 	&	0.148 	&	0.70 	&	0	&	2.364 	&	2.374 	&	0.43 &\,\cite{Piekarewicz2002} 	\\
FAMB4	&	37.01 	&	107.85 	&	275.06 	&	-263.80 	&	-16.00 	&	0.148 	&	0.70 	&	0	&	2.364 	&	2.386 	&	0.94 &\,\cite{Piekarewicz2002} 	\\
FAMB5	&	37.01 	&	107.66 	&	300.06 	&	-190.23 	&	-16.00 	&	0.148 	&	0.70 	&	0	&	2.364 	&	2.398 	&	1.43 &\,\cite{Piekarewicz2002} 	\\
FAMC1	&	28.01 	&	81.40 	&	200.03 	&	-593.39 	&	-16.00 	&	0.148 	&	0.70 	&	0	&	2.364 	&	2.363 	&	-0.04 &\,\cite{Piekarewicz2002} 	\\
FAMC2	&	28.01 	&	81.21 	&	225.05 	&	-465.00 	&	-16.00 	&	0.148 	&	0.70 	&	0	&	2.364 	&	2.373 	&	0.40 &\,\cite{Piekarewicz2002} 	\\
FAMC3	&	28.01 	&	81.03 	&	250.05 	&	-355.30 	&	-16.00 	&	0.148 	&	0.70 	&	0	&	2.364 	&	2.385 	&	0.89 &\,\cite{Piekarewicz2002} 	\\
FAMC4	&	28.00 	&	80.82 	&	275.06 	&	-263.80 	&	-16.00 	&	0.148 	&	0.70 	&	0	&	2.364 	&	2.397 	&	1.40 &\,\cite{Piekarewicz2002} 	\\
FAMC5	&	28.00 	&	80.63 	&	300.06 	&	-190.23 	&	-16.00 	&	0.148 	&	0.70 	&	0	&	2.364 	&	2.409 	&	1.89 &\,\cite{Piekarewicz2002} 	\\
GL1	&	32.49 	&	94.67 	&	199.97 	&	-619.56 	&	-16.30 	&	0.153 	&	0.70 	&	0	&	2.321 	&	2.317 	&	-0.20 	&\,\cite{compactstars} \\
GL2	&	32.51 	&	91.56 	&	200.04 	&	-608.36 	&	-16.30 	&	0.153 	&	0.75 	&	0	&	2.137 	&	2.088 	&	-2.35 	&\,\cite{compactstars} \\
GL3	&	32.51 	&	89.07 	&	200.07 	&	-580.52 	&	-16.31 	&	0.153 	&	0.80 	&	0	&	1.953 	&	1.871 	&	-4.42 	&\,\cite{compactstars}\\
GL4	&	32.51 	&	94.36 	&	250.07 	&	-380.98 	&	-16.31 	&	0.153 	&	0.70 	&	0	&	2.321 	&	2.339 	&	0.74 	&\,\cite{compactstars}\\
GL5	&	32.51 	&	91.21 	&	250.11 	&	-455.08 	&	-16.31 	&	0.153 	&	0.75 	&	0	&	2.137 	&	2.137 	&	0.01 	&\,\cite{compactstars}\\
GL6	&	32.51 	&	88.78 	&	250.06 	&	-547.18 	&	-16.31 	&	0.153 	&	0.80 	&	0	&	1.953 	&	1.959 	&	0.28 	&\,\cite{compactstars}\\
GL7	&	32.51 	&	90.90 	&	300.09 	&	-384.18 	&	-16.31 	&	0.153 	&	0.75 	&	0	&	2.137 	&	2.183 	&	2.11 	&\,\cite{compactstars}\\
GL8	&	32.51 	&	88.48 	&	300.07 	&	-609.94 	&	-16.31 	&	0.153 	&	0.80 	&	0	&	1.953 	&	2.006 	&	2.64 	&\,\cite{compactstars}\\
GL82	&	36.23 	&	101.48 	&	285.74 	&	-438.15 	&	-16.00 	&	0.153 	&	0.77 	&	0	&	2.064 	&	2.157 	&	4.31 &\,\cite{Glendenning1982}	\\
GL9	&	32.51 	&	89.93 	&	210.08 	&	-564.97 	&	-16.31 	&	0.153 	&	0.78 	&	0	&	2.027 	&	1.972 	&	-2.78  &\,\cite{compactstars}	\\
GM1	&	32.52 	&	94.04 	&	300.50 	&	-215.72 	&	-16.33 	&	0.153 	&	0.70 	&	0	&	2.321 	&	2.363 	&	1.76  &\,\cite{Glendenning1991}	\\
GM2	&	32.51 	&	89.38 	&	300.01 	&	-505.58 	&	-16.31 	&	0.153 	&	0.78 	&	0	&	2.027 	&	2.078 	&	2.47 	 &\,\cite{Glendenning1991}	\\
GM3	&	32.51 	&	89.75 	&	240.04 	&	-512.13 	&	-16.30 	&	0.153 	&	0.78 	&	0	&	2.027 	&	2.018 	&	-0.43  &\,\cite{Glendenning1991}		\\
GPS	&	32.49 	&	88.59 	&	299.59 	&	-590.30 	&	-15.96 	&	0.150 	&	0.80 	&	0	&	1.979 	&	2.028 	&	2.44 	&\,\cite{Ghosh1995}	\\
Hybrid	&	37.30 	&	118.07 	&	230.01 	&	-108.39 	&	-16.24 	&	0.148 	&	0.60 	&	0	&	2.732 	&	2.756 	&	0.86 &\,\cite{Piekarewicz2009} 	\\
MS2	&	35.00 	&	111.02 	&	249.92 	&	78.13 	&	-15.75 	&	0.148 	&	0.60 	&	0	&	2.732 	&	2.761 	&	1.06 &\,\cite{Muller1996}	\\
NL06	&	39.33 	&	124.51 	&	195.09 	&	-346.54 	&	-16.05 	&	0.147 	&	0.60 	&	0	&	2.740 	&	2.761 	&	0.73 &\,\cite{Dutra2014}	\\
NL065	&	38.98 	&	117.76 	&	256.87 	&	-220.20 	&	-16.37 	&	0.150 	&	0.65 	&	0	&	2.531 	&	2.553 	&	0.86 & \,\cite{Dutra2014}	\\
NL07	&	38.52 	&	112.27 	&	276.45 	&	-296.52 	&	-16.49 	&	0.150 	&	0.70 	&	0	&	2.347 	&	2.366 	&	0.83 &\,\cite{Dutra2014}	\\
NL075	&	38.96 	&	110.44 	&	281.12 	&	-419.88 	&	-16.64 	&	0.151 	&	0.75 	&	0	&	2.154 	&	2.173 	&	0.87 &\,\cite{Dutra2014}	\\
NL1	&	43.46 	&	140.07 	&	211.09 	&	-32.69 	&	-16.42 	&	0.152 	&	0.57 	&	0	&	2.808 	&	2.810 	&	0.06 &\,\cite{Reinhard1989}	\\
NL1J4	&	40.00 	&	130.08 	&	211.70 	&	4.30 	&	-16.42 	&	0.152 	&	0.57 	&	0	&	2.808 	&	2.819 	&	0.39 &\,\cite{Centelles1998}	 	\\
NL1J5	&	50.00 	&	160.08 	&	211.70 	&	4.30 	&	-16.42 	&	0.152 	&	0.57 	&	0	&	2.808 	&	2.809 	&	0.02  &\,\cite{Centelles1998}		\\
NL2	&	43.86 	&	129.66 	&	399.37 	&	68.42 	&	-17.03 	&	0.146 	&	0.67 	&	0	&	2.491 	&	2.550 	&	2.31 &\,\cite{Reinhard1989}	\\
NL3	&	37.40 	&	118.53 	&	271.53 	&	202.91 	&	-16.24 	&	0.148 	&	0.60 	&	0	&	2.748 	&	2.780 	&	1.14 &\cite{Lalazissis1997}	\\
NL3-II	&	37.70 	&	120.03 	&	271.72 	&	247.88 	&	-16.26 	&	0.149 	&	0.59 	&	0	&	2.760 	&	2.785 	&	0.89 &\,\cite{Lalazissis1997}	 	\\
NL3*	&	38.68 	&	123.10 	&	258.25 	&	156.36 	&	-16.31 	&	0.150 	&	0.59 	&	0	&	2.752 	&	2.773 	&	0.78 &\cite{Lalazissis2009}	\\
NL4	&	36.24 	&	114.51 	&	270.34 	&	162.95 	&	-16.16 	&	0.148 	&	0.60 	&	0	&	2.732 	&	2.761 	&	1.05 	&\,\cite{Bozena2004}\\
NLB	&	35.01 	&	108.35 	&	421.02 	&	737.40 	&	-15.77 	&	0.148 	&	0.61 	&	0	&	2.695 	&	2.749 	&	1.98 &\,\cite{Serot1997}	\\
NLB1	&	33.04 	&	102.50 	&	280.44 	&	108.58 	&	-15.79 	&	0.162 	&	0.62 	&	0	&	2.539 	&	2.575 	&	1.40  &\,\cite{Reinhard1989}	\\
NLB2	&	33.10 	&	112.01 	&	245.58 	&	615.63 	&	-15.79 	&	0.163 	&	0.55 	&	0	&	2.789 	&	2.795 	&	0.24  &\,\cite{Reinhard1989}	\\
NLC	&	35.02 	&	108.12 	&	224.46 	&	-271.56 	&	-15.77 	&	0.148 	&	0.63 	&	0	&	2.621 	&	2.647 	&	0.99  &\,\cite{Serot1997}	\\
NLD	&	35.01 	&	101.36 	&	343.21 	&	-87.36 	&	-15.77 	&	0.148 	&	0.70 	&	0	&	2.364 	&	2.421 	&	2.38 	&\cite{Serot1992} \\
NLM	&	30.00 	&	87.02 	&	200.00 	&	-600.49 	&	-16.00 	&	0.160 	&	0.70 	&	0	&	2.262 	&	2.269 	&	0.30 &\,\cite{Centelles1998}		\\
NLM2	&	30.00 	&	86.95 	&	200.00 	&	-675.54 	&	-17.00 	&	0.160 	&	0.70 	&	0	&	2.262 	&	2.264 	&	0.08 &\,\cite{Centelles1998}	 	\\
NLM3	&	30.00 	&	87.45 	&	200.00 	&	591.82 	&	-16.00 	&	0.145 	&	0.70 	&	0	&	2.389 	&	2.385 	&	-0.17 &\,\cite{Centelles1998}		\\
NLM4	&	30.00 	&	86.25 	&	300.00 	&	-196.02 	&	-16.00 	&	0.160 	&	0.70 	&	0	&	2.262 	&	2.314 	&	2.26 &\,\cite{Centelles1998}		\\
NLM5	&	30.00 	&	103.18 	&	200.00 	&	216.98 	&	-16.00 	&	0.160 	&	0.55 	&	0	&	2.814 	&	2.822 	&	0.29 	&\,\cite{Centelles1998}	\\
NLM6	&	40.00 	&	117.02 	&	200.00 	&	-600.49 	&	-16.00 	&	0.160 	&	0.70 	&	0	&	2.262 	&	2.258 	&	-0.17 &\,\cite{Centelles1998}		\\
NLRA	&	38.90 	&	118.73 	&	320.47 	&	191.18 	&	-16.25 	&	0.157 	&	0.63 	&	0	&	2.545 	&	2.580 	&	1.37 &\,\cite{Chung2000} 	\\
NLRA1	&	36.45 	&	115.00 	&	285.23 	&	251.33 	&	-16.15 	&	0.147 	&	0.60 	&	0	&	2.740 	&	2.772 	&	1.13 &\cite{Rashdan2001}	\\
NL$\rho$	&	30.37 	&	84.61 	&	240.76 	&	-464.67 	&	-16.05 	&	0.160 	&	0.75 	&	0	&	2.078 	&	2.089 	&	0.54 &\,\cite{Liu2002}	\\
NLS	&	42.07 	&	132.03 	&	262.94 	&	88.05 	&	-16.44 	&	0.150 	&	0.60 	&	0	&	2.715 	&	2.734 	&	0.71 &\,\cite{Reinhard1988}	\\
NLSH	&	36.13 	&	113.37 	&	355.64 	&	573.02 	&	-16.36 	&	0.146 	&	0.60 	&	0	&	2.749 	&	2.790 	&	1.47 &\cite{Lalazissis1997}	\\
NL-VT1	&	39.03 	&	123.75 	&	179.03 	&	-482.13 	&	-16.09 	&	0.150 	&	0.60 	&	0	&	2.715 	&	2.732 	&	0.64  &\,\cite{Bender1999}		\\
NLZ	&	41.72 	&	134.33 	&	172.84 	&	-393.69 	&	-16.18 	&	0.151 	&	0.58 	&	0	&	2.780 	&	2.732 	&	-1.75 &\,\cite{Bender1999}	\\
NLZ2	&	39.01 	&	126.22 	&	172.23 	&	-385.09 	&	-16.06 	&	0.151 	&	0.58 	&	0	&	2.780 	&	2.794 	&	0.50 &\,\cite{Bender1999}		\\
P-067	&	41.07 	&	122.37 	&	231.63 	&	-412.46 	&	-16.31 	&	0.157 	&	0.67 	&	0	&	2.398 	&	2.410 	&	0.52 &\,\cite{Sulaksono2005}	\\
P-070	&	41.85 	&	122.11 	&	245.07 	&	-402.05 	&	-16.25 	&	0.163 	&	0.70 	&	0	&	2.236 	&	2.252 	&	0.71 &\,\cite{Sulaksono2005}		\\
P-075	&	42.95 	&	121.59 	&	271.29 	&	-436.17 	&	-16.51 	&	0.173 	&	0.75 	&	0	&	1.968 	&	2.018 	&	2.51 &\,\cite{Sulaksono2005}		\\
P-080	&	39.63 	&	109.68 	&	259.93 	&	-536.00 	&	-15.84 	&	0.162 	&	0.80 	&	0	&	1.877 	&	1.909 	&	1.70 &\,\cite{Sulaksono2005}		\\
Q1	&	36.44 	&	115.39 	&	241.86 	&	-12.29 	&	-16.10 	&	0.148 	&	0.60 	&	0	&	2.732 	&	2.758 	&	0.95 &\,\cite{Furnstahl1997}	\\
RMF301	&	32.50 	&	89.63 	&	253.86 	&	-499.15 	&	-16.30 	&	0.153 	&	0.78 	&	0	&	2.027 	&	2.035 	&	0.39 &\,\cite{Dadi2010}	\\
RMF302	&	32.50 	&	89.66 	&	249.71 	&	-502.32 	&	-16.30 	&	0.153 	&	0.78 	&	0	&	2.027 	&	2.030 	&	0.16 	&\,\cite{Dadi2010}\\
RMF303	&	32.50 	&	89.66 	&	248.88 	&	-503.03 	&	-16.30 	&	0.153 	&	0.78 	&	0	&	2.027 	&	2.029 	&	0.11 &\,\cite{Dadi2010}	\\
RMF304	&	32.50 	&	89.67 	&	248.04 	&	-503.77 	&	-16.30 	&	0.153 	&	0.78 	&	0	&	2.027 	&	2.028 	&	0.06 &\,\cite{Dadi2010}	\\
RMF305	&	32.50 	&	89.68 	&	246.37 	&	-505.32 	&	-16.30 	&	0.153 	&	0.78 	&	0	&	2.027 	&	2.026 	&	-0.04 &\,\cite{Dadi2010}	\\
RMF306	&	32.50 	&	89.22 	&	244.69 	&	-524.52 	&	-16.30 	&	0.153 	&	0.79 	&	0	&	1.990 	&	1.988 	&	-0.10 &\,\cite{Dadi2010}	\\
RMF307	&	32.50 	&	89.23 	&	243.84 	&	-524.95 	&	-16.30 	&	0.153 	&	0.79 	&	0	&	1.990 	&	1.987 	&	-0.15 &\,\cite{Dadi2010}	\\
RMF308	&	32.50 	&	89.23 	&	242.99 	&	-525.40 	&	-16.30 	&	0.153 	&	0.79 	&	0	&	1.990 	&	1.986 	&	-0.21 &\,\cite{Dadi2010}	\\
RMF309	&	32.50 	&	89.24 	&	241.30 	&	-526.37 	&	-16.30 	&	0.153 	&	0.79 	&	0	&	1.990 	&	1.984 	&	-0.32 	&\,\cite{Dadi2010}\\
RMF310	&	32.50 	&	89.26 	&	238.75 	&	-528.04 	&	-16.30 	&	0.153 	&	0.79 	&	0	&	1.990 	&	1.980 	&	-0.49 &\,\cite{Dadi2010}	\\
RMF311	&	32.50 	&	89.26 	&	237.89 	&	-528.66 	&	-16.30 	&	0.153 	&	0.79 	&	0	&	1.990 	&	1.979 	&	-0.55 &\,\cite{Dadi2010}	\\
RMF312	&	32.50 	&	89.27 	&	237.03 	&	-529.31 	&	-16.30 	&	0.153 	&	0.79 	&	0	&	1.990 	&	1.978 	&	-0.60 &\,\cite{Dadi2010}	\\
RMF313	&	32.50 	&	88.83 	&	235.31 	&	-546.31 	&	-16.30 	&	0.153 	&	0.80 	&	0	&	1.953 	&	1.940 	&	-0.68 &\,\cite{Dadi2010}	\\
RMF314	&	32.50 	&	88.84 	&	234.43 	&	-546.55 	&	-16.30 	&	0.153 	&	0.80 	&	0	&	1.953 	&	1.939 	&	-0.74 	&\,\cite{Dadi2010}\\
RMF315	&	32.50 	&	88.84 	&	234.01 	&	-546.67 	&	-16.30 	&	0.153 	&	0.80 	&	0	&	1.953 	&	1.938 	&	-0.78 &\,\cite{Dadi2010}	\\
RMF316	&	32.50 	&	88.84 	&	233.57 	&	-546.81 	&	-16.30 	&	0.153 	&	0.80 	&	0	&	1.953 	&	1.937 	&	-0.81 &\,\cite{Dadi2010}	\\
RMF317	&	32.50 	&	88.85 	&	232.70 	&	-547.11 	&	-16.30 	&	0.153 	&	0.80 	&	0	&	1.953 	&	1.936 	&	-0.87 &\,\cite{Dadi2010}	\\
RMF401	&	32.50 	&	93.79 	&	229.99 	&	-477.83 	&	-16.30 	&	0.153 	&	0.71 	&	0	&	2.284 	&	2.288 	&	0.14 &\,\cite{Dadi2010}	\\
RMF402	&	32.50 	&	93.77 	&	231.99 	&	-469.22 	&	-16.30 	&	0.153 	&	0.71 	&	0	&	2.284 	&	2.289 	&	0.18 	&\,\cite{Dadi2010}\\
RMF403	&	32.50 	&	93.13 	&	229.99 	&	-486.52 	&	-16.30 	&	0.153 	&	0.72 	&	0	&	2.248 	&	2.246 	&	-0.08 &\,\cite{Dadi2010}	\\
RMF404	&	32.50 	&	93.11 	&	231.99 	&	-478.60 	&	-16.30 	&	0.153 	&	0.72 	&	0	&	2.248 	&	2.247 	&	-0.03 &\,\cite{Dadi2010}	\\
RMF405	&	32.50 	&	93.10 	&	233.99 	&	-470.79 	&	-16.30 	&	0.153 	&	0.72 	&	0	&	2.248 	&	2.248 	&	0.03 &\,\cite{Dadi2010}	\\
RMF406	&	32.50 	&	89.76 	&	233.99 	&	-520.01 	&	-16.30 	&	0.153 	&	0.78 	&	0	&	2.027 	&	2.010 	&	-0.83 &\,\cite{Dadi2010}	\\
RMF407	&	32.50 	&	92.50 	&	229.99 	&	-493.80 	&	-16.30 	&	0.153 	&	0.73 	&	0	&	2.211 	&	2.204 	&	-0.31 	&\,\cite{Dadi2010}\\
RMF408	&	32.50 	&	92.49 	&	231.99 	&	-486.54 	&	-16.30 	&	0.153 	&	0.73 	&	0	&	2.211 	&	2.205 	&	-0.24 &\,\cite{Dadi2010}	\\
RMF409	&	32.50 	&	92.47 	&	233.99 	&	-479.42 	&	-16.30 	&	0.153 	&	0.73 	&	0	&	2.211 	&	2.207 	&	-0.17 &\,\cite{Dadi2010}	\\
RMF410	&	32.50 	&	92.49 	&	235.99 	&	-472.42 	&	-16.30 	&	0.153 	&	0.73 	&	0	&	2.211 	&	2.209 	&	-0.10 &\,\cite{Dadi2010}	\\
RMF411	&	32.50 	&	91.90 	&	229.99 	&	-500.08 	&	-16.30 	&	0.153 	&	0.74 	&	0	&	2.174 	&	2.162 	&	-0.54 	&\,\cite{Dadi2010}\\
RMF412	&	32.50 	&	91.89 	&	231.99 	&	-493.56 	&	-16.30 	&	0.153 	&	0.74 	&	0	&	2.174 	&	2.164 	&	-0.45 	&\,\cite{Dadi2010}\\
RMF413	&	32.50 	&	91.88 	&	233.99 	&	-487.17 	&	-16.30 	&	0.153 	&	0.74 	&	0	&	2.174 	&	2.166 	&	-0.37 &\,\cite{Dadi2010}	\\
RMF414	&	32.50 	&	91.86 	&	235.99 	&	-480.90 	&	-16.30 	&	0.153 	&	0.74 	&	0	&	2.174 	&	2.168 	&	-0.29 &\,\cite{Dadi2010}	\\
RMF415	&	32.50 	&	91.33 	&	229.98 	&	-505.93 	&	-16.30 	&	0.153 	&	0.75 	&	0	&	2.137 	&	2.121 	&	-0.75 &\,\cite{Dadi2010}	\\
RMF416	&	32.50 	&	91.32 	&	231.98 	&	-500.18 	&	-16.30 	&	0.153 	&	0.75 	&	0	&	2.137 	&	2.123 	&	-0.65 &\,\cite{Dadi2010}	\\
RMF417	&	32.50 	&	91.31 	&	233.99 	&	-494.54 	&	-16.30 	&	0.153 	&	0.75 	&	0	&	2.137 	&	2.126 	&	-0.55 &\,\cite{Dadi2010}	\\
RMF418	&	32.50 	&	91.29 	&	235.98 	&	-489.09 	&	-16.30 	&	0.153 	&	0.75 	&	0	&	2.137 	&	2.128 	&	-0.45 &\,\cite{Dadi2010}	\\
RMF419	&	32.50 	&	90.79 	&	229.99 	&	-511.76 	&	-16.30 	&	0.153 	&	0.76 	&	0	&	2.100 	&	2.081 	&	-0.92 &\,\cite{Dadi2010}	\\
RMF420	&	32.50 	&	90.78 	&	231.99 	&	-506.84 	&	-16.30 	&	0.153 	&	0.76 	&	0	&	2.100 	&	2.084 	&	-0.80 &\,\cite{Dadi2010}	\\
RMF421	&	32.50 	&	90.76 	&	233.99 	&	-502.07 	&	-16.30 	&	0.153 	&	0.76 	&	0	&	2.100 	&	2.086 	&	-0.69 &\,\cite{Dadi2010}	\\
RMF422	&	32.50 	&	90.27 	&	229.99 	&	-518.23 	&	-16.30 	&	0.153 	&	0.77 	&	0	&	2.064 	&	2.042 	&	-1.05 &\,\cite{Dadi2010}	\\
RMF423	&	32.50 	&	90.26 	&	231.99 	&	-514.22 	&	-16.30 	&	0.153 	&	0.77 	&	0	&	2.064 	&	2.045 	&	-0.91 &\,\cite{Dadi2010}	\\
RMF424	&	32.50 	&	89.22 	&	245.99 	&	-523.93 	&	-16.30 	&	0.153 	&	0.79 	&	0	&	1.990 	&	1.990 	&	-0.02 &\,\cite{Dadi2010}	\\
RMF425	&	32.50 	&	89.20 	&	247.99 	&	-523.14 	&	-16.30 	&	0.153 	&	0.79 	&	0	&	1.990 	&	1.992 	&	0.11 &\,\cite{Dadi2010}	\\
RMF426	&	32.50 	&	89.19 	&	249.99 	&	-522.50 	&	-16.30 	&	0.153 	&	0.79 	&	0	&	1.990 	&	1.994 	&	0.23 &\,\cite{Dadi2010}	\\
RMF427	&	32.50 	&	88.83 	&	235.98 	&	-546.14 	&	-16.30 	&	0.153 	&	0.80 	&	0	&	1.953 	&	1.941 	&	-0.63&\,\cite{Dadi2010} 	\\
RMF428	&	32.50 	&	88.82 	&	237.98 	&	-545.76 	&	-16.30 	&	0.153 	&	0.80 	&	0	&	1.953 	&	1.944 	&	-0.49 &\,\cite{Dadi2010}	\\
RMF429	&	32.50 	&	88.81 	&	239.99 	&	-545.53 	&	-16.30 	&	0.153 	&	0.80 	&	0	&	1.953 	&	1.946 	&	-0.35 &\,\cite{Dadi2010}	\\
RMF430	&	32.50 	&	88.79 	&	241.99 	&	-545.46 	&	-16.30 	&	0.153 	&	0.80 	&	0	&	1.953 	&	1.949 	&	-0.22 &\,\cite{Dadi2010}	\\
RMF431	&	32.50 	&	88.78 	&	243.98 	&	-545.54 	&	-16.30 	&	0.153 	&	0.80 	&	0	&	1.953 	&	1.951 	&	-0.09 &\,\cite{Dadi2010}	\\
RMF432	&	32.50 	&	88.77 	&	245.95 	&	-545.78 	&	-16.30 	&	0.153 	&	0.80 	&	0	&	1.953 	&	1.954 	&	0.04 &\,\cite{Dadi2010}	\\
RMF433	&	32.50 	&	88.76 	&	247.99 	&	-546.17 	&	-16.30 	&	0.153 	&	0.80 	&	0	&	1.953 	&	1.956 	&	0.16 &\,\cite{Dadi2010}	\\
RMF434	&	32.50 	&	88.75 	&	249.99 	&	-546.72 	&	-16.30 	&	0.153 	&	0.80 	&	0	&	1.953 	&	1.959 	&	0.28 &\,\cite{Dadi2010}	\\
RSk*	&	30.03 	&	81.70 	&	216.60 	&	-530.22 	&	-15.77 	&	0.160 	&	0.79 	&	0	&	1.931 	&	1.909 	&	-1.12 &\,\cite{Centelles1998}	\\
S271	&	36.46 	&	106.21 	&	271.00 	&	-295.54 	&	-16.24 	&	0.148 	&	0.70 	&	0	&	2.360 	&	2.380 	&	0.83 &\cite{Horowitz2002}	\\
S271v1	&	35.74 	&	95.93 	&	271.00 	&	-295.54 	&	-16.24 	&	0.148 	&	0.70 	&	0	&	2.360 	&	2.380 	&	0.84 &\cite{Horowitz2002}	\\
S271v2	&	35.06 	&	86.87 	&	271.00 	&	-295.54 	&	-16.24 	&	0.148 	&	0.70 	&	0	&	2.360 	&	2.340 	&	-0.87 &\cite{Horowitz2002}	\\
S271v3	&	34.42 	&	78.86 	&	271.00 	&	-295.54 	&	-16.24 	&	0.148 	&	0.70 	&	0	&	2.360 	&	2.338 	&	-0.96&\cite{Horowitz2002} 	\\
S271v4	&	33.83 	&	71.76 	&	271.00 	&	-295.54 	&	-16.24 	&	0.148 	&	0.70 	&	0	&	2.360 	&	2.341 	&	-0.83 &\cite{Horowitz2002}	\\
S271v5	&	33.27 	&	65.44 	&	271.00 	&	-295.54 	&	-16.24 	&	0.148 	&	0.70 	&	0	&	2.360 	&	2.344 	&	-0.69 &\cite{Horowitz2002}	\\
S271v6	&	32.74 	&	59.81 	&	271.00 	&	-295.54 	&	-16.24 	&	0.148 	&	0.70 	&	0	&	2.360 	&	2.347 	&	-0.56 &\cite{Horowitz2002}	\\
SRK3M5	&	23.49 	&	82.44 	&	299.86 	&	966.01 	&	-16.00 	&	0.150 	&	0.55 	&	0	&	2.899 	&	2.926 	&	0.91 &\,\cite{Centelles1992}	\\
SRK3M7	&	28.73 	&	79.70 	&	299.95 	&	-364.09 	&	-16.00 	&	0.150 	&	0.75 	&	0	&	2.163 	&	2.211 	&	2.18 &\,\cite{Centelles1992} \\
VT	&	39.72 	&	127.08 	&	172.74 	&	-468.05 	&	-16.09 	&	0.153 	&	0.59 	&	0	&	2.726 	&	2.740 	&	0.50 	&\cite{Rufa1988}\\
IOPB-I	&	33.36 	&	63.70 	&	222.65 	&	-116.44 	&	-16.11 	&	0.149 	&	0.60 	&	0.002907	&	2.136 	&	2.150 	&	0.66 &\cite{Kumar2018} 	\\
FSUGarnet	&	30.92 	&	50.96 	&	229.63 	&	8.47 	&	-16.23 	&	0.153 	&	0.58 	&	0.003916	&	2.047 	&	2.070 	&	1.09  &\,\cite{Chen2015}	\\
IU-FSU	&	31.30 	&	47.21 	&	231.33 	&	-290.15 	&	-16.40 	&	0.155 	&	0.61 	&	0.005	&	1.960 	&	1.940 	&	-1.03 &\,\cite{Fattoyev2010}	\\
IU-FSU*	&	29.85 	&	50.30 	&	235.69 	&	-259.40 	&	-16.02 	&	0.150 	&	0.61 	&	0.005	&	1.978 	&	1.960 	&	-0.94 	&\,\cite{Agrawal2012}\\
FSU-J0	&	33.45 	&	68.14 	&	229.20 	&	-322.00 	&	-16.31 	&	0.148 	&	0.61 	&	0.005283	&	1.977 	&	1.930 	&	-2.46 &\,\cite{FanLi2022}	\\
SRV00	&	33.49 	&	65.23 	&	223.94 	&	-224.08 	&	-16.11 	&	0.149 	&	0.61 	&	0.003526	&	2.053 	&	2.040 	&	-0.64  &\,\cite{VirenderThakur2022} 	\\
TM1	&	36.89 	&	110.80 	&	281.16 	&	-285.23 	&	-16.26 	&	0.145 	&	0.63 	&	0.002816	&	2.097 	&	2.180 	&	3.81 &\,\cite{Sugahara1994}	\\
OMEG1	&	35.06 	&	70.00 	&	256.00 	&	-300.62 	&	-16.38 	&	0.148 	&	0.62 	&	0.003617	&	2.028 	&	2.130 	&	4.80 &\,\cite{Miyatsu2023}	\\
OMEG2	&	33.00 	&	45.00 	&	256.00 	&	-300.56 	&	-16.38 	&	0.148 	&	0.62 	&	0.003617	&	2.028 	&	2.070 	&	2.04 &\,\cite{Miyatsu2023}	\\
OMEG3	&	30.00 	&	20.00 	&	256.00 	&	-300.28 	&	-16.38 	&	0.148 	&	0.62 	&	0.003617   &	2.028 	&	2.070 	&	2.04 &\,\cite{Miyatsu2023}	\\
SRV01	&	33.75 	&	63.82 	&	221.78 	&	-192.94 	&	-16.11 	&	0.149 	&	0.60 	&	0.003416	&	2.069 	&	2.060 	&	-0.43 	 &\,\cite{VirenderThakur2022}\\
SRV02	&	33.31 	&	61.49 	&	222.05 	&	-197.82 	&	-16.09 	&	0.149 	&	0.60 	&	0.003361	&	2.073 	&	2.070 	&	-0.14  &\,\cite{VirenderThakur2022}	\\
SRV03	&	33.54 	&	58.06 	&	221.72 	&	-188.44 	&	-16.12 	&	0.149 	&	0.60 	&	0.003628	&	2.051 	&	2.080 	&	1.38  &\,\cite{VirenderThakur2022} 	\\
SRV04	&	33.34 	&	55.31 	&	221.11 	&	-178.12 	&	-16.11 	&	0.149 	&	0.60 	&	0.003481	&	2.066 	&	2.130 	&	3.02 &\,\cite{VirenderThakur2022}	\\
FSU-d6.7	&	32.75 	&	53.50 	&	229.20 	&	-322.00 	&	-16.31 	&	0.148 	&	0.61 	&	0.005283	&	1.977 	&	2.050 	&	3.54  &\,\cite{FanLi2022}	\\
FSU-d6.2	&	32.53 	&	48.21 	&	229.20 	&	-322.00 	&	-16.31 	&	0.148 	&	0.61 	&	0.005283	&	1.977 	&	2.100 	&	5.84 &\,\cite{FanLi2022}	\\
\hline
\end{longtable*}

%

\end{document}